\documentclass{aa}

\usepackage[dvipsnames]{xcolor}
\usepackage{xspace}
\usepackage{amsmath}
\usepackage{rotating}
\usepackage{threeparttable}
\usepackage{enumerate}
\usepackage{color}
\usepackage{graphicx}
\usepackage{soul}
\usepackage{ulem}[normalem]
\usepackage{txfonts}
\usepackage{hyperref}

\newcommand{\hstfull}{Hubble Space Telescope\xspace}
\newcommand{\hst}{HST\xspace}

\newcommand{\jwst}{JWST\xspace}
\newcommand{\gaia}{\textit{Gaia}\xspace}

\newcommand{\radxs}{\texttt{RADXS}\xspace}

\newcommand{\jpass}{\texttt{jwst1pass}\xspace}
\newcommand{\kstwo}{\texttt{KS2}\xspace}
\newcommand{\masyr}{mas yr$^{-1}$\xspace}

\defcitealias{2024BedinNIRCam6397}{Paper~I}
\defcitealias{2024RomanBDs6397}{Paper~II}

\begin{document}

\title{\jwst imaging of the closest globular clusters\,--\,IV.\\ Chemistry, luminosity, and mass functions of the lowest-mass members in the NIRISS parallel fields\thanks{Astro-photometric catalogs and stacked images are available at the CDS via anonymous ftp to \href{cdsarc.u-strasbg.fr}{cdsarc.u-strasbg.fr} (130.79.128.5) or via \href{http://cdsweb.u-strasbg.fr/cgi-bin/qcat?J/A+A/}{http://cdsweb.u-strasbg.fr/cgi-bin/qcat?J/A+A/}.}}

\titlerunning{\jwst Imaging of the Closest Globular Clusters\,--\,IV.}

\author{M. Libralato
    \inst{1,2}
    \and
    R. Gerasimov
    \inst{3}
    \and
    L. Bedin
    \inst{1}
    \and
    J. Anderson
    \inst{4}
    \and
    D. Apai
    \inst{5,6}
    \and
    A. Bellini
    \inst{4}
    \and
    A. J. Burgasser
    \inst{7}
    \and
    M. Griggio
    \inst{1,4,8}
    \and
    D. Nardiello
    \inst{9}
    \and
    M. Salaris
    \inst{10}
    \and
    M. Scalco
    \inst{1}
    \and
    E. Vesperini
    \inst{11}
    }

\institute{
INAF - Osservatorio Astronomico di Padova, Vicolo dell'Osservatorio 5, Padova I-35122, Italy \\
\email{mattia.libralato@inaf.it}
\and
AURA for the European Space Agency, Space Telescope Science Institute, 3700 San Martin Drive, Baltimore, MD 21218, USA
\and
Department of Physics and Astronomy, University of Notre Dame, Nieuwland Science Hall, Notre Dame, IN 46556, USA
\and
Space Telescope Science Institute, 3700 San Martin Drive, Baltimore, MD 21218, USA
\and
Department of Astronomy and Steward Observatory, The University of Arizona, 933 N. Cherry Avenue, Tucson, AZ 85721, USA
\and
Lunar and Planetary Laboratory, The University of Arizona, 1629 E. University Blvd., Tucson, AZ 85721, USA
\and
Department of Astronomy \& Astrophysics, University of California, San Diego, La Jolla, California 92093, USA
\and
Dipartimento di Fisica e Scienze della Terra, Università di Ferrara, Via Giuseppe Saragat 1, Ferrara I-44122, Italy
\and
Dipartimento di Fisica e Astronomia ``Galileo Galilei'', Universit{\`a} di Padova, Vicolo dell'Osservatorio 3, Padova I-35122, Italy
\and
Astrophysics Research Institute, Liverpool John Moores University, 146 Brownlow Hill, Liverpool L3 5RF, UK
\and
Department of Astronomy, Indiana University, Swain West, 727 E. 3rd Street, Bloomington, IN 47405, USA
}

\date{Received 27 June 2024; Accepted 04 September 2024}
 
\abstract
{
We present observations of the two closest globular clusters, NGC~6121 and NGC~6397, taken with the NIRISS detector of \jwst. The combination of our new \jwst data with archival \hstfull (\hst) images allows us to compute proper motions, disentangle cluster members from field objects, and probe the main sequence (MS) of the clusters down to $<$0.1$M_\odot$ as well as the brighter part of the white-dwarf sequence. We show that theoretical isochrones fall short in modeling the low-mass MS and discuss possible explanations for the observed discrepancies. Our analysis suggests that the lowest-mass members of both clusters are significantly more metal-rich and oxygen-poor than their higher-mass counterparts. It is unclear whether the difference is caused by a genuine mass-dependent chemical heterogeneity, low-temperature atmospheric processes altering the observed abundances, or systematic shortcomings in the models. We computed the present-day local luminosity and mass functions of the two clusters; our data reveal a strong flattening of the mass function indicative of a significant preferential loss of low-mass stars in agreement with previous dynamical models for these two clusters. We have made our NIRISS astro-photometric catalogs and stacked images publicly available to the community.
}


\maketitle

\section{Introduction}\label{sec:intro}

Globular clusters (GCs) have been the subject of countless investigations on a wide range of topics because they are ideal laboratories for testing models of star formation and evolution. While the most massive stars in GCs have been extensively studied in recent decades, too little is known about the least massive members along the main sequence (MS) and, in particular in the regime of brown dwarfs. Very-faint objects among very-bright stars in GCs have always posed a challenge even to the \hstfull (\hst), requiring a significant amount of telescope time to be observed, if they could be observed at all \citep[]{harvey_6397,BD_hunt_1,2019MNRAS.486.2254D}. \jwst \citep{2023GardnerJWST} has already shown that objects with masses $<$0.1 M$_\odot$ and brown dwarfs are within its reach \citep[e.g.,][]{2023MNRAS.521L..39N,2024arXiv240106681M}.

The scientific interest in photometric observations of the lowest-mass GC members is predominantly three-fold. First, the colors of low-mass stars and brown dwarfs are sensitive to the abundances of key elements, including oxygen, titanium, carbon and alkali metals \citep{clouds_and_chemistry,roman_cs21,roman_47tuc}. The sensitivity primarily arises from the extreme pressure broadening of atomic lines and dominant molecular opacity. Photometric abundances measured near the end of the MS may be compared to the spectroscopic observations of upper-MS and post-MS members to constrain the magnitude of chemical variations with stellar mass \citep{NGC_6752_spread,NGC_6752_individual,2012ApJ...745...27M,michele_6752,JWST_low_MS_phot_1,JWST_low_MS_phot_3,MP_in_6440}. The presence or absence of such variations has been suggested as a means to differentiate between the proposed origins of the presently unexplained chemical scatter among GC members \citep[the so-called multiple populations; see reviews in][]{2015RenzinimPOPs,mPOPs_review_1,2020CassimPOPsRv,mPOPs_review_3}. In addition, GCs provide a unique opportunity to study the discrepancy between primordial and observed element abundances in cool stars and brown dwarfs, driven by low-temperature phenomena such as condensation and gravitational settling of dust, and nonequilibrium chemistry. This effect is likely responsible for the disagreement between theoretical and observed brown dwarf luminosities near the hydrogen-burning limit \citep{2024arXiv240106681M}, and inconsistent abundances in star or brown-dwarf binary systems \citep{BD_inconsistent_chemistry}.

Second, the luminosity functions (LFs) of brown dwarfs and low-mass stars just above the hydrogen-burning limit are sensitive to the age of the parent population, as the energy production by nuclear fusion in the cores of these objects is insufficient to attain energy equilibrium \citep{HBL_1,HBL_2,adam_gap}. GCs are prime candidates for this dating technique, because their old ages \citep[$\gtrsim 12\ \mathrm{Gyr}$,][]{GC_ages,GC_ages_2} and large sizes \citep[$\gtrsim 10^5$ members, ][]{GC_masses} ensure that the stellar-substellar transition in the LF is both prominent and well populated \citep{roman_omegacen,roman_47tuc}.

Third, the low-mass regime of the present-day GC mass functions (MFs) provides an important constraint on the effects of dynamical evolution, which remain the largest source of uncertainty in the estimates of the initial MF in GCs. At present, inferred MFs of GCs are largely consistent with a universal initial MF, as evident from the strong correlation between the measured MFs, central densities and dynamical relaxation times \citep{MF_vs_environment,sollima_1}, the lack of clear environmental dependence among less dynamically evolved clusters \citep{NGC6397_MF_2}, and direct comparison to simulations \citep{harvey_6397,IMF_from_MF,sollima_2}. Including the lowest-mass members in MF studies not only extends the mass baseline, but also potentially probes a distinct regime of star formation (e.g., where additional formation channels for substellar objects --such as disk or filament fragmentation-- need to be considered). While no clear discontinuities in the substellar MF of star clusters have been observed so far (see review in \citealt{IMF_clusters_review}), their existence is tentatively suggested by the binary properties of brown dwarfs in the field \citep{IMF_anomaly_1,IMF_anomaly_2,IMF_anomaly_3,IMF_anomaly_4}.

The present paper is the fourth of a series aimed at investigating the coolest sources in the two closest GCs: NGC~6121 and NGC~6397 \citep[Paper\,I]{2024BedinNIRCam6397}. The first three papers of the series focused on NGC~6397, specifically, \citetalias{2024BedinNIRCam6397} focused on the white dwarfs (WDs), \citet[hereafter Paper~II]{2024RomanBDs6397} on the brown dwarfs, and \citet[Paper~III]{2024ScalcoJWST} on the multiple populations. Here, we analyze the faint-end of the MS using the \jwst's Near Infrared Imager and Slitless Spectrograph (NIRISS) camera \citep{2012SPIE.8442E..2RD}. In Section~\ref{sec:data}, we describe the data reduction in detail, including the tweaks applied to the official \jwst pipeline to improve the final astrometric and photometric products. In Section~\ref{sec:analysis}, we discuss how we combined \hst and \jwst data to compute proper motions (PMs) and present color-magnitude diagrams (CMDs) of the two clusters. We compare our photometry for the two clusters with a set of isochrones in Section~\ref{sec:iso}. Finally, the LFs and MFs for NGC~6121 and NGC~6397 in our fields are presented in Sect.~\ref{sec:lfmf}.

\begin{figure*}
    \centering
    \includegraphics[height=7.05cm]{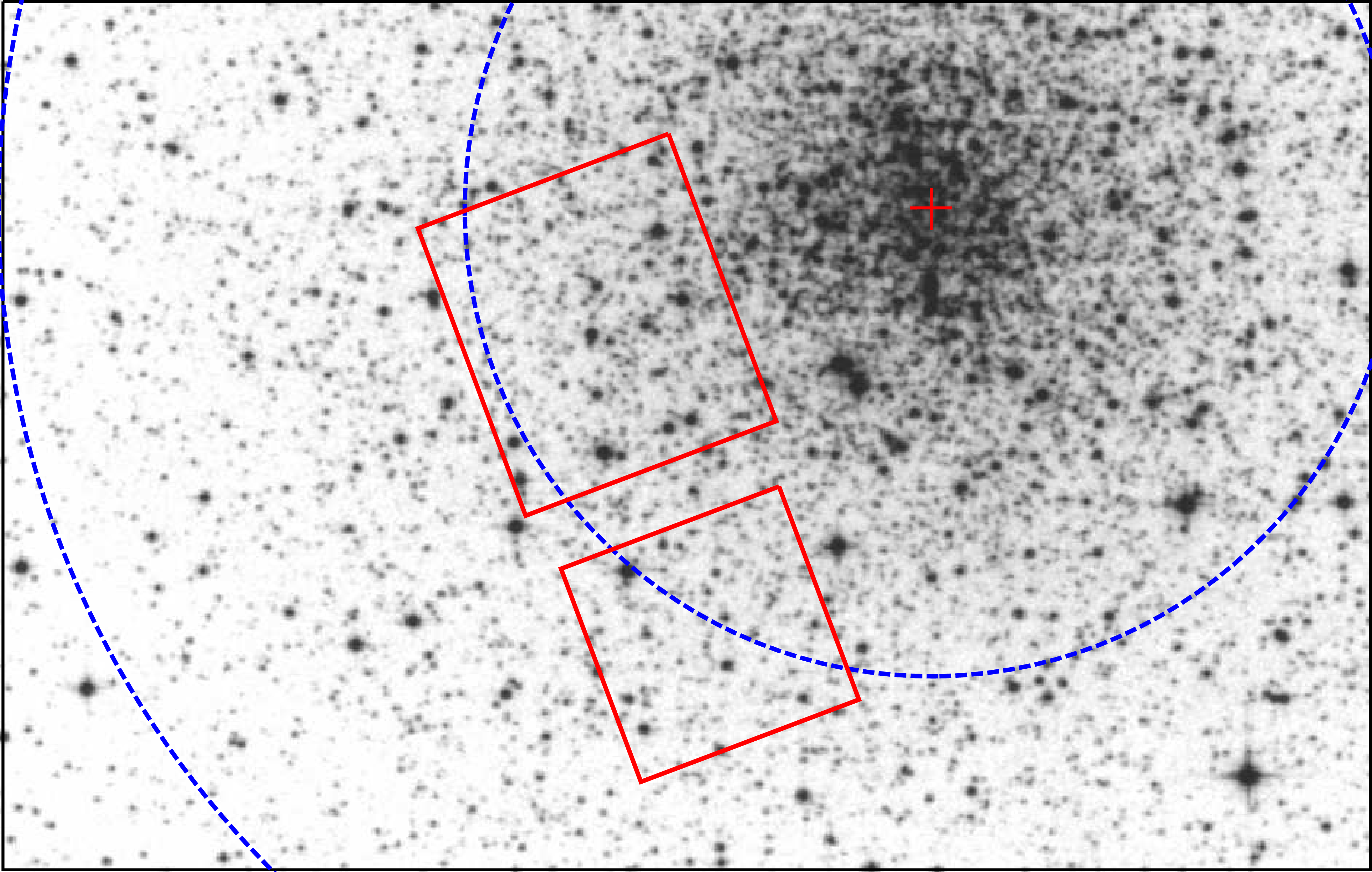}
    \includegraphics[height=7.06cm]{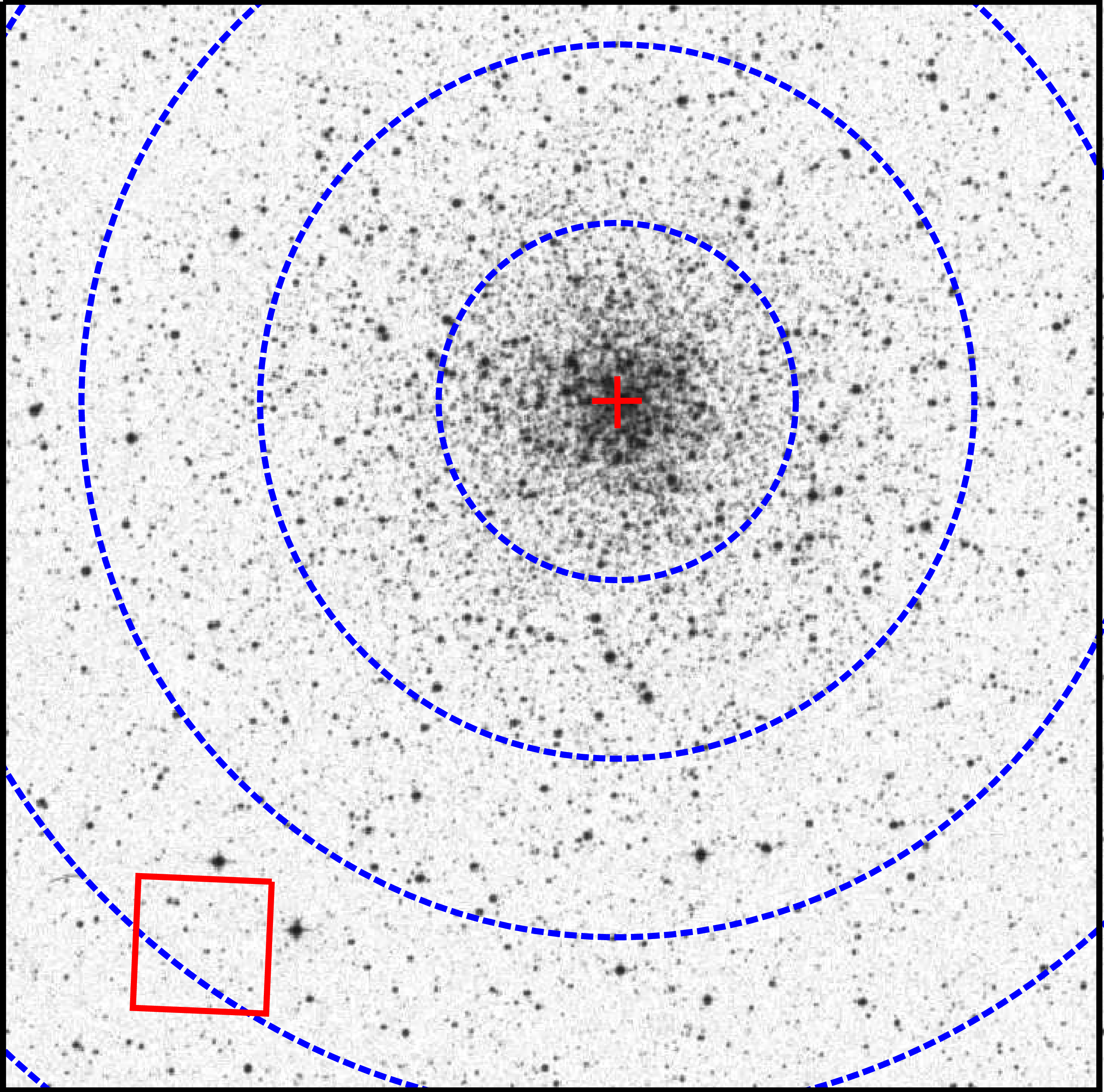}
    \includegraphics[width=\textwidth]{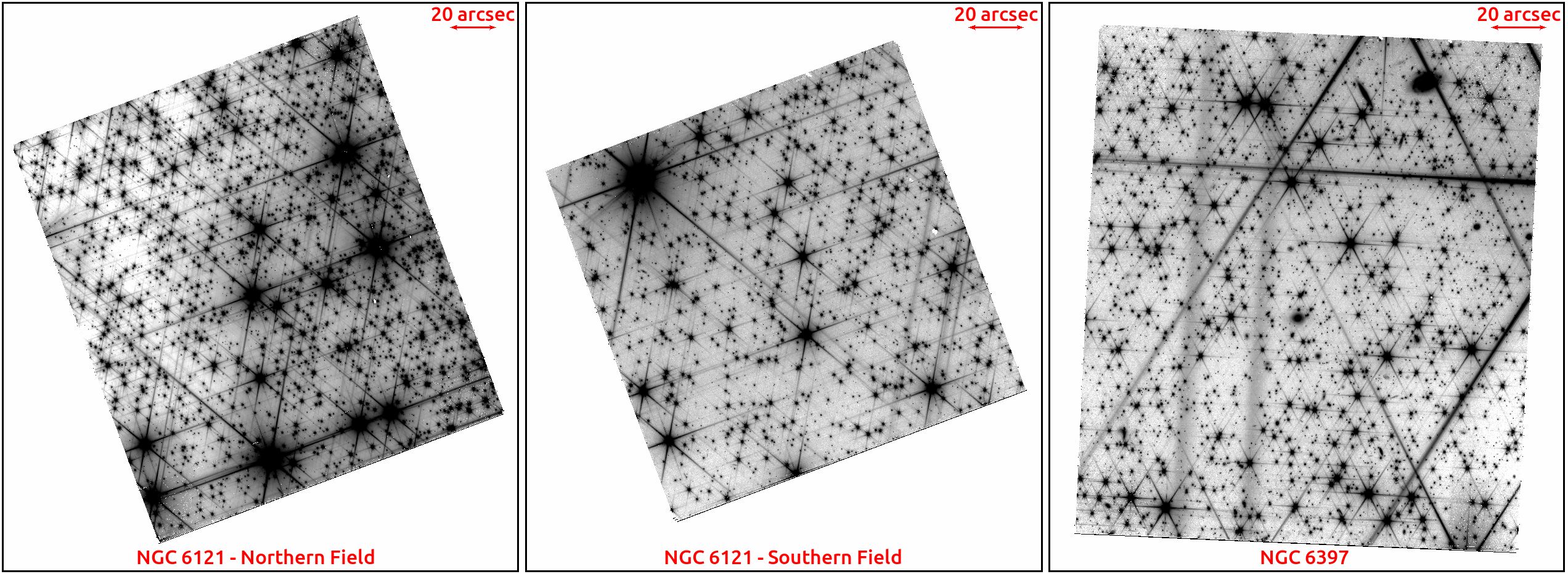}
    \caption{Analyzed field of view. (Top): Gray-scale infrared images of NGC~6121 (left) and NGC~6397 (right) from the Digital Sky Survey 2. The centers of the clusters (taken from the Baumgardt GC database) are highlighted with a red cross. The blue dashed circles have radii equal to $N$ times the half-light radius, with $N = [1,2]$ for NGC~6121 and [1,2,3,4,5] for NGC~6397. The red polygons in each panel highlight the coverage of the NIRISS data. (Bottom): Gray-scale stacked images of NGC~6121 and NGC~6397 made with the GO-1979 NIRISS data. In all panels, north is up and east is to the left.}
    \label{fig:fov}
\end{figure*}

\section{Data sets and reduction}\label{sec:data}

NGC~6121 and NGC~6397 were observed with \jwst as part of program GO-1979 (PI: Bedin) using the Near Infrared Camera \citep[NIRCam,][]{2023RiekeNIRCam} in primary and NIRISS in parallel. The main science goal of the program is to reach the faintest sources of these GCs in the field of view of the primary instrument. Therefore, the choices of telescope pointing, dither pattern and the observing setup are driven by the requirements of NIRCam. Figure~\ref{fig:fov} illustrates the position of our NIRISS fields with respect to the center of each cluster. Below we described the data sets and reduction procedures.

\subsection{NGC~6121 data set}\label{sec:ngc6121data}

NGC~6121 was observed by \jwst on 2023 April 9-10 in two different fields located at about the half-light radius ($r_{\rm h}$$\sim$4.65 arcmin according to the GC database of Holger Baumgardt\footnote{\href{https://people.smp.uq.edu.au/HolgerBaumgardt/globular/}{https://people.smp.uq.edu.au/HolgerBaumgardt/globular/}}) from the center of the cluster (always taken from the Baumgardt database), which we refer to as the ``Northern'' and ``Southern'' fields.

The Northern field is centered at about (RA,Dec.) $=$ (245.958201,$-26.544989$) deg, which is $\sim$3.6 arcmin from the center of the cluster. The field was observed with 12 F150W-filter images in a 2$\times$3 dither pattern (two subdithers per position) for a total field of view of 3.0$\times$2.7 arcmin$^2$. Exposures were taken with the NIS readout, 11 groups, and one integration, for an effective exposure time of 472.418\,s for each of the 12 exposures.

The Southern field is centered at about (RA,Dec.) $=$ (245.937502,$-26.596299$) and is located at $\sim$4.8 arcmin from the center of NGC~6121. Only three exposures with small dithers were taken (field of view of 2.2$\times$2.2 arcmin$^2$). The observing setup consisted again in F150W images with the NISRAPID readout, 11 groups and one integration (effective exposure time of 118.104 s).

Both NIRISS fields overlap (the Northern partially and the Southern completely) with existing \hst data from the GO-12911 program (PI: Bedin) acquired between 2013 January 18 and May 19. This data set contains 30$\times$180s exposures taken with the Wide Field Camera (WFC) of the Advanced Camera for Surveys (ACS) in F606W and F814W filters.

\subsection{NGC~6397 data set}\label{sec:ngc6397data}

The NIRISS observations of NGC~6397 are centered at about (RA,Dec.) $=$ (265.372698,$-53.826364$) deg ($\sim$11.5 arcmin away from the center of the cluster, i.e., 4$r_{\rm h}$, assuming $r_{\rm h}$$\sim$3 arcmin as per the Baumgardt database). The data were taken on 2023 March 14 and comprise 12 F150W-filter images organized in a 2$\times$3 dither pattern (with two images per dither position), covering a field of view of 2.2$\times$2.2 arcmin$^2$. Each image was obtained with the NIS readout mode, 14 groups and one integration per exposure, for an effective exposure time of 601.259\,s.

Analogously to the case of NGC~6121, this NIRISS field partially overlaps with existing \hst data from GO-9480 (PI: Rhodes) taken on 2003 March 27--28. This data set is made up of five ACS/WFC images (3$\times$ 700\,s, 1$\times$ 558\,s, 1$\times$ 400\,s) in the F775W filter.

\subsection{\jwst data reduction}\label{sec:jwstred}

We downloaded the NIRISS level-1b, uncalibrated (\texttt{\_uncal}) fits files from the Mikulski Archive for Space Telescopes (MAST)\footnote{\href{https://mast.stsci.edu/search/ui/}{https://mast.stsci.edu/search/ui/}.}. We processed these images using the official \jwst pipeline\footnote{\href{https://github.com/spacetelescope/jwst}{https://github.com/spacetelescope/jwst}.} version 1.12.5 \citep{2023BushouseJWSTpipeline} through the stages 1 and 2 to obtain level-2b (\texttt{\_cal}) images. We ran the pipeline using the default values for all but two steps.

First, we turned on the charge-migration step in the stage-1 pipeline and set the signal-threshold parameter of the step to 25\,000. This choice allowed us to improve the photometry and astrometry for the brightest stars near saturation, which are severely affected by the brighter-fatter effect \citep[e.g.,][]{2023LibralatoNIRISS}. Second, we let the ramp-fit step to fit the ramps of saturated pixels using the first frame of each integration (``suppress\_one\_group'' $=$ False), which, in turn, increased the dynamical range of our data and improved measurements of saturated stars in our images. The mitigation of the brighter-fatter effect and the recovery of saturated stars were essential steps in our work because they improved the quality of the sources that are included in the \gaia Early Data Release 3 (EDR3) catalog \citep{2016GaiaCit,2021GaiaEDR3}, which we used to set a common reference frame system.  

We initially used the publicly available FORTRAN code \jpass\footnote{\href{https://www.stsci.edu/~jayander/JWST1PASS/CODE/JWST1PASS/}{https://www.stsci.edu/$\sim$jayander/JWST1PASS/CODE/JWST1PASS/}.} \citep[Anderson et al., in preparation]{2023LibralatoNIRISS,2024LibralatoMIRI} to measure positions and fluxes of bright stars in each NIRISS exposure by fitting effective point-spread functions (ePSFs). The library NIRISS ePSFs were tailored for each image\footnote{In each image, we found bright, unsaturated stars and measured their position and flux using the library ePSF models. We then subtracted a model of each star based on the parameters obtained from the fit. The residuals of the subtraction were collected in a grid, the size of which varies from a 1$\times$1 to a 5$\times$5 depending on the number of sources at disposal, and then averaged. Finally, the ePSF models were modified according to the residual grid (a linear interpolation was used to obtain the ePSF perturbation at any given location). The entire process was iterated nine times.}. Positions were corrected for geometric distortion. Library ePSFs and geometric-distortion corrections were described and made publicly available for the NIRISS imager by \citet{2023LibralatoNIRISS}.

We cross-matched bright stars in each NIRISS catalog with those in the \gaia EDR3 catalog. The \gaia catalog (projected onto a tangent plane centered at about the center of our NIRISS fields after positions were moved to the 2023 epoch by means of the \gaia PMs) was used to setup a common reference frame system. We imposed that the $x$ and $y$ axes point towards west and north, respectively, and fixed the pixel scale to 40 mas pixel$^{-1}$ to somewhat better sample the image stacks. We iteratively cross-identified the same stars in all images, and applied six-parameter linear transformations\footnote{Almost all sources in the NIRISS data in common with the \gaia EDR3 catalog are saturated. Thus, we used saturated stars for the initial registration on the the \gaia reference frame. Once the master frame was set up, we only used bright, unsaturated stars in the process.} to transform stellar positions in each NIRISS catalog onto the master frame. The NIRISS photometry was registered to that of a NIRISS single-image catalog. Once on the same reference system, positions and instrumental magnitudes were averaged to create a preliminary astro-photometric catalog.

The final stage of the data reduction was obtained using a NIRISS-tailored version of the software \kstwo \citep[Anderson et al., in preparation; see also][]{2017BelliniwCenI,2022LibralatoPMcat}. \kstwo performs the so-called second-pass photometry, which makes use of all images at once to detect objects too faint to be measured in a single image. Also, \kstwo measures all sources after all close-by neighbors have been ePSF subtracted from the image. The astro-photometric catalogs made by \kstwo contain several diagnostic parameters that can be used to select a sample of well-measured stars. We refer the reader to \citet{2022LibralatoPMcat} for a complete description of the outputs of \kstwo.

\kstwo does not measure saturated stars, for which positions and fluxes are taken from the output of the first-pass photometry \citep[see discussion in][]{2017BelliniwCenI}. Saturated and very-bright stars are used by \kstwo to construct appropriate masks around them to prevent the software from detecting PSF artifacts. In our fields, there are many very bright stars for which modeling some PSF features is a complex task; also, these stars can extend for thousands of pixels (Fig.~\ref{fig:fov}). For this reason, we made use of the \texttt{Python} package WebbPSF \citep{2012PerrinWebbPSF,2014PerrinWebbPSF}, which can provide simulated PSFs for all \jwst imagers taking into account models of the telescope and optical state of an instrument but not detector effects. A single WebbPSF PSF of 3072$\times$3072 NIRISS pixel$^2$ was placed at the position and rescaled by the flux of each saturated/very bright star (which is similar fashion to what is done during the ePSF-fitting stage) to make these masks. Although not perfect, this step allowed us to reject most of the PSF artifacts around bright objects.

\begin{figure*}[th!]
    \centering
    \includegraphics[height=10.5cm, keepaspectratio]{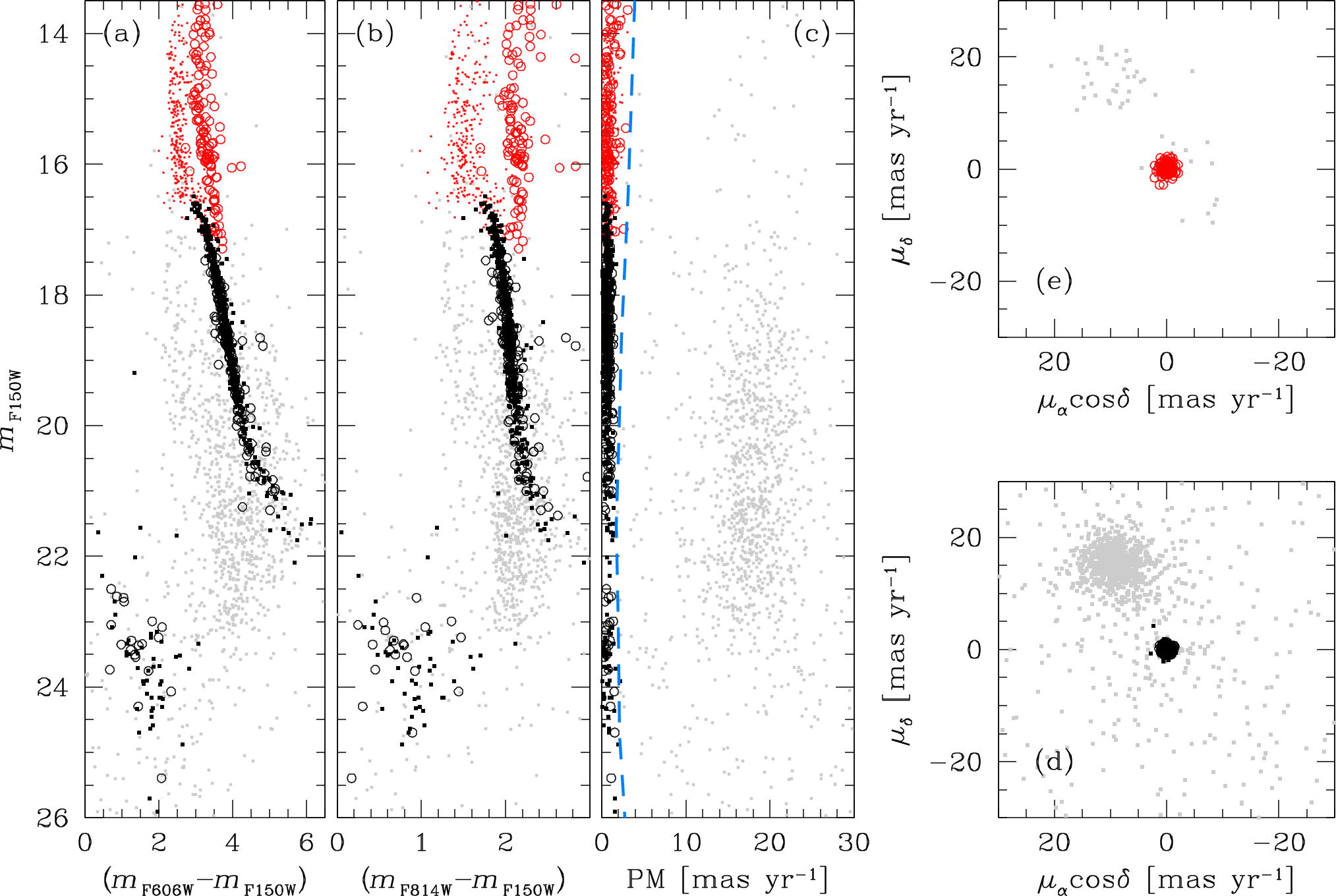}
    \caption{Overview results for NGC~6121. Panels (a) and (b) present the $m_{\rm F150W}$ as a function of $(m_{\rm F606W}-m_{\rm F150W})$ and $(m_{\rm F814W}-m_{\rm F150W})$ CMDs, respectively. We plot $m_{\rm F150W}$ as a function of PM in panel (c). In all these panels, gray points represent field stars, while all other colors refer to cluster members according to their PMs. Black open circles and dots are unsaturated stars measured in the Northern and Southern fields, respectively. The corresponding symbols in red mark saturated sources in NIRISS. The blue dashed line in panel (c) was used to infer cluster membership. Panels (d) and (e) show the VPD for unsaturated and saturated stars, respectively.}
    \label{fig:m4overview}
\end{figure*}

\begin{figure*}
    \centering
    \includegraphics[height=10.5cm, keepaspectratio]{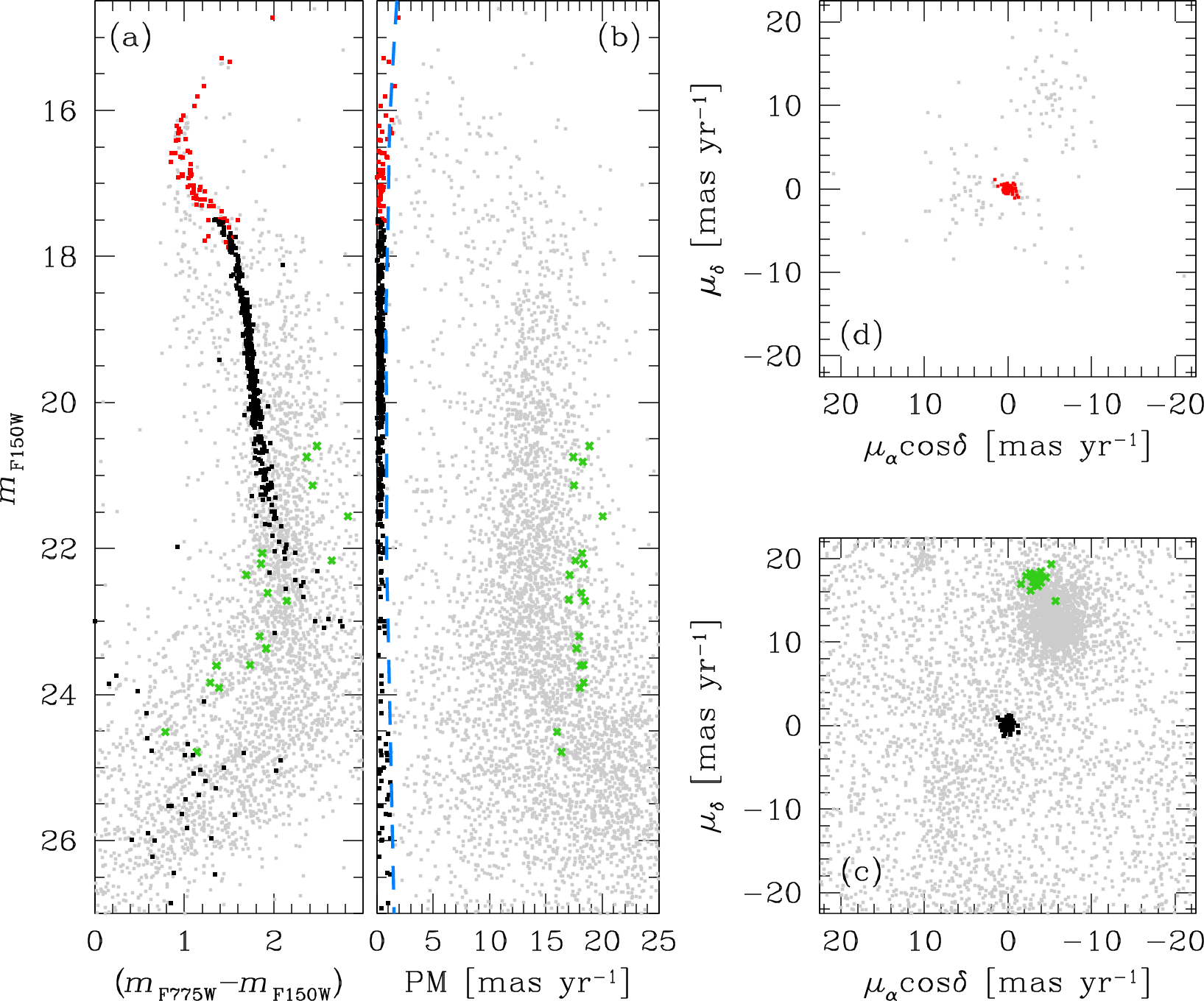}
    \caption{Similar to Fig.~\ref{fig:m4overview} but for NGC~6397. The green crosses in panels (a), (b), and (c) refer to a sample of background galaxies selected from the NIRISS stacked image (see Appendix~\ref{appendix:abspm} for details). The clump of points in panel (c) at about ($\mu_\alpha \cos\delta$,$\mu_\delta$) $\sim$ (10,20) \masyr is caused by mismatches between residual PSF artifacts in the NIRISS data and real field stars in the \hst catalog that moved by 11 pixels over 20 years.}
    \label{fig:ngc6397overview}
\end{figure*}

The photometric calibration of our NIRISS catalogs was performed as follows. First, we used the stage-3 pipeline of \jwst to combine all \texttt{\_cal} fits files of each field and make a level-3 (\texttt{\_i2d}) mosaic image. We then measured the flux of bright and unsaturated stars in this \texttt{\_i2d} image via aperture photometry using as aperture radius that enclosing the 70\% of the total energy of a point source. The sky was estimated in an annulus region with inner and outer radii enclosing the 80\% and 85\% of the total energy, respectively. These radii are provided by the aperture-correction reference file\footnote{Reference file: ``jwst\_niriss\_apcorr\_008.fits''.} of NIRISS in the \jwst Calibration Reference Data System (CRDS). We chose these values because they allow us to directly use the official NIRISS aperture corrections to correct our photometry for the finite aperture. We converted the aperture-corrected surface brightness (the unit of the \texttt{\_i2d} images is MJy sr$^{-1}$) of each star to a flux density in Jansky using the conversion factor provided in the fits header and then defined the magnitudes in the VEGAmag system as:
\begin{displaymath}
    \begin{split}
  m_{\rm VEGA} &= -2.5 \times \log ({\rm flux \, density}) + 8.9 + {\rm ZP}_{\rm VEGA} \\
  &= m_{\rm AB} + {\rm ZP}_{\rm VEGA} \phantom{1} ,
  \end{split}
\end{displaymath}
where ZP$_{\rm VEGA}$ is the AB-to-VEGA zero-point available in CRDS\footnote{Reference file: ``jwst\_niriss\_abvegaoffset\_0003.asdf''.}. Finally, we cross-identified the same stars in our \kstwo-based and aperture-based catalogs and computed a zero-point to add to the \kstwo instrumental magnitudes to put them in the calibrated VEGAmag system. These zero-points are 33.184 and 31.684 for the Northern- and Southern-field photometry of NGC~6121, respectively, and 33.447 for the photometry of NGC~6397.

\subsection{\hst data reduction}\label{sec:hstred}

The \hst data reduction was performed on \texttt{\_flc} images \citep[unresampled exposures that are dark and bias corrected, flat-fielded, and pipeline-corrected for the charge-transfer-efficiency defects as described in][]{2018acs..rept....4A} through a combination of first- and second-pass photometric stages as previously described for \jwst. We made use of publicly available library ePSFs and geometric-distortion corrections\footnote{\href{https://www.stsci.edu/~jayander/HST1PASS/}{https://www.stsci.edu/$\sim$jayander/HST1PASS/}.} \citep{2006acs..rept....1A}. The master frame was again setup by means of the \gaia EDR3 catalog after positions were moved to the epoch of the \hst observations.

The calibration of the \kstwo-based \hst instrumental magnitudes was obtained using the \texttt{\_drc} (for ACS/WFC or WFC3/UVIS) images, the official aperture corrections and VEGA-mag zero-points\footnote{See the resources provided here: \href{https://www.stsci.edu/hst/instrumentation/acs/data-analysis}{https://www.stsci.edu/hst/instrumentation/acs/data-analysis}, \href{https://www.stsci.edu/hst/instrumentation/wfc3/data-analysis/photometric-calibration}{https://www.stsci.edu/hst/instrumentation/wfc3/data-analysis/photometric-calibration}.} \citep[see][]{2017BelliniwCenI}.

\section{A first look at the CMDs}\label{sec:analysis}

For each cluster and field, the astro-photometric catalogs made with \hst and \jwst data were combined by cross-identifying the same stars in both catalogs. Positions as measured in the \hst catalog were transformed onto the reference frame of the \jwst catalog by means of six-parameter linear transformations. The coefficients of these transformations were computed using only bright, unsaturated cluster stars (iteratively identified through their location on the CMD and their positional displacement). Relative PMs were defined as the positional displacements between epochs divided by the average temporal baseline and multiplied by the pixel scale of the reference frame (40 mas pixel$^{-1}$). Because we selected cluster stars as a reference, our PMs are computed relative to the bulk motion of cluster. This means that in the vector-point diagram (VPD), the distribution of the cluster members is centered at (0,0) by construction, and all other objects are placed in a region of the VPD according to their motion relative to the cluster. In the following, the photometry used in all CMDs has been corrected for systematic errors using artificial-star tests as described in Appendix~\ref{sec:artstar}.

\subsection{NGC~6121}\label{sec:firstlook6121}

The \hst-\jwst CMDs of NGC~6121 are shown in Fig.~\ref{fig:m4overview} (panels a and b). Photometry is corrected for the effects of the differential reddening as in \citet{2012A&A...540A..16M} and \citet{2017BelliniwCenDR}, assuming E($B-V$)$=$0.37 from \citet{2012HendricksM4red}. Black markers represent unsaturated cluster members, while saturated stars are plotted in red. Open circles and dots refer to objects measured in the Northern and Southern NIRISS fields, respectively. The Northern and Southern fields have different read-out modes (NIS and NISRAPID, respectively) and therefore different behavior for saturated objects, which explains the discrepancy between the two cluster sequences in red. All other objects are plotted as gray points. The cluster membership is inferred by hand in the plot PM as a function of magnitude (blue dashed line in panel b of Fig.~\ref{fig:m4overview}). This threshold is defined as the compromise between excluding field stars with cluster-like motion and including genuine cluster members. To take into account the larger PM errors of saturated and very faint stars, the membership criteria are less severe for these objects.

The photometry for saturated stars is still not reliable (red points in Fig.~\ref{fig:m4overview}), as it creates unphysical blue/red turns in the CMDs, and is different for the two sets of NIRISS data. On the other hand, the PMs of saturated and unsaturated objects are in sufficient agreement with each other and are precise enough to assess the cluster membership (see panels d and e). These pieces of information might suggest that brighter-fatter effects have a more severe impact on the photometry of very bright sources rather than on their astrometry, at least for NIRISS.

The CMDs in Fig.~\ref{fig:m4overview} are corrected for a small zero-point ($\sim$0.05 mag) between the NIRISS photometry of unsaturated stars using the Northern- and Southern-field data. We do not know the reason for this offset, but we have corrected for it by applying the offset to the Southern-field data taken with the NISRAPID readout. We chose to correct the photometry of this data set because for both NGC~6121 and NGC~6397 we noticed a better agreement between our CMDs with NIRISS data taken with the NIS readout and theoretical isochrones (Sect.~\ref{sec:iso}). While differential reddening could also explain the discrepancy in the photometry of the two fields \citep[the extinction in NGC~6121 varies significantly across the field, with $\Delta$E($B-V$)$\sim$0.2;][]{2012HendricksM4red}, it is unlikely to be the cause, because this zero-point is not present in the F606W and F814W photometry.

The combination of \hst and \jwst data allows us to probe the MS of NGC~6121 down to $m_{\rm F150W} \sim 22$ ($m_{\rm F606W} \sim 27.5$). This depth is comparable with what was obtained by \citet{2009BedinM4end}, who used F606W (and F775W) observations for their \hst program GO-10146 that were about ten times longer (1200\,s). However, we fall $\sim$1 mag short with respect to \citeauthor{2009BedinM4end} along the WD sequence.
    
\subsection{NGC~6397}

A CMD of NGC~6397 constructed with $m_{\rm F150W}$ as a function of $(m_{\rm F775W}-m_{\rm F150W})$ and corrected for differential reddening using E($B-V$)$=$0.22 \citep{2018ApJ...864..147C} is presented in panel (a) of Fig.~\ref{fig:ngc6397overview}. As noted for NGC~6121, the astrometry of saturated stars seems consistent with that of unsaturated objects, whereas there is a clear systematic issue with the saturated-star photometry. We also note that a sample of background galaxies (green crosses in Fig.~\ref{fig:ngc6397overview}) was measured in both epochs. We used these galaxies to measure the absolute PM of the clusterm and the results are presented in Appendix~\ref{appendix:abspm}.

As for NGC~6121, we can measure stars along the MS of NGC~6397 down to $m_{\rm F150W} \sim 23$. We can also probe the brightest part of the WD sequence, but we are again limited by the shallow \hst data (see Appendix~\ref{sec:artstar}).

\section{Isochrone fitting}\label{sec:iso}

The following subsections describe the isochrone-fitting procedure for the two clusters. Hereafter, we consider only well-measured cluster members, that is, those objects that have: (i) a \radxs value \citep[excess or deficiency of flux outside the core of the star compared to what was predicted by the ePSF;][]{2008ApJ...678.1279B} within $\pm$0.1; (ii) a fractional flux within the fitting radius prior to neighbor subtraction \citep[e.g.,][]{2022LibralatoPMcat} lower than 0.25; (iii) a flux at least 3$\sigma$ above the local sky; (iv) a rejection rate (number of images where a source was measured over the number of images in which a source was found) of lower than 40\%; and (v) a PM measurement. All CMDs presented in our work are corrected for the so-called input--output effect described in Appendix~\ref{sec:artstar}.

\subsection{NGC~6121}

\begin{figure}
    \centering
    \includegraphics[width=\columnwidth]{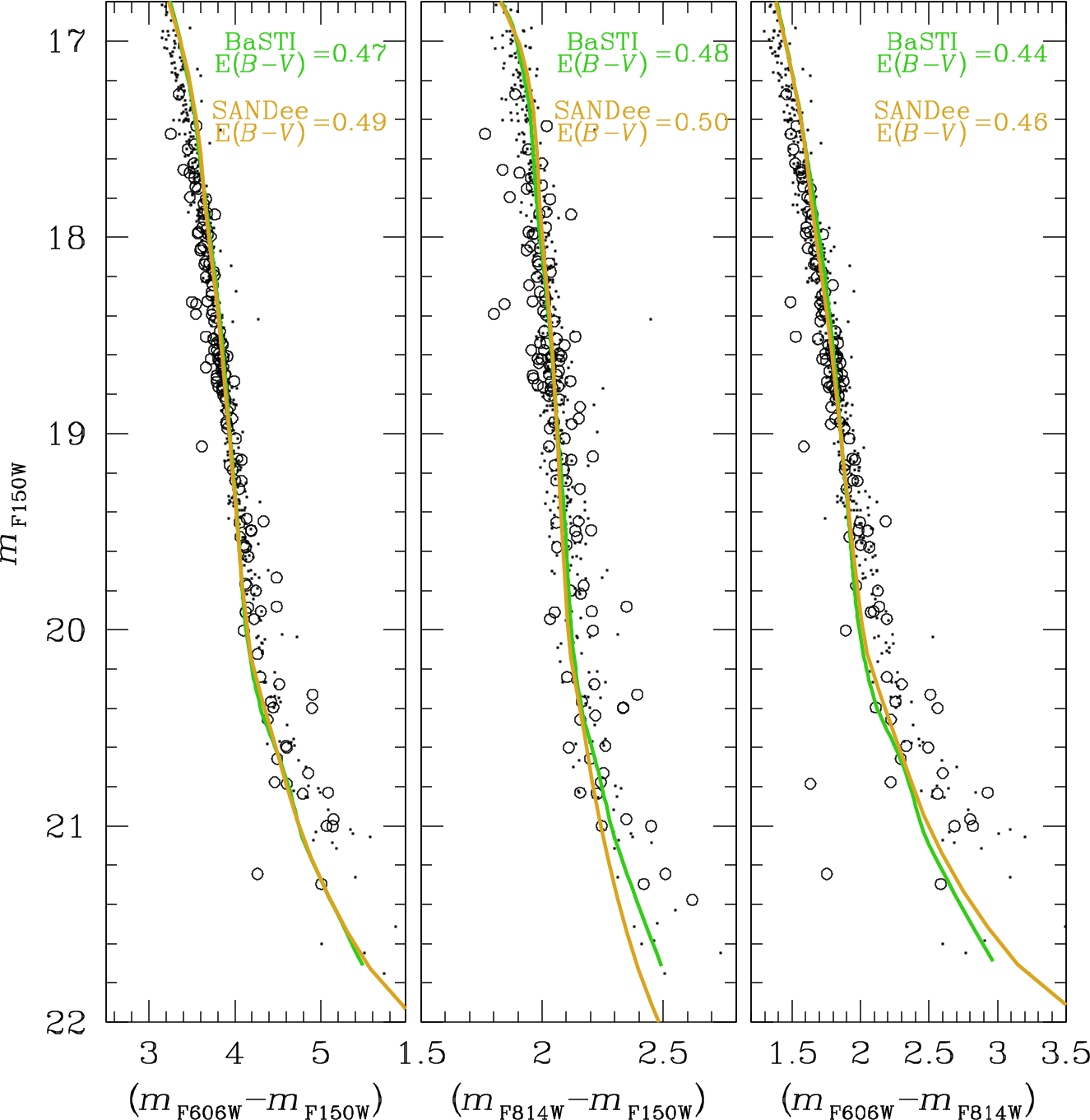}
    \caption{Collection of CMDs for the MS of NGC~6121. Only well-measured cluster members are shown using the same symbols as in Fig.~\ref{fig:m4overview}. The green and yellow lines represent the best-fit BaSTI and SANDee isochrones, respectively, using distance and metallicity from the literature. The best-fit extinction for each isochrone is provided in each panel.}
    \label{fig:m4iso1}
\end{figure}

We compared our CMDs of NGC~6121 with model isochrones from the ``a Bag of Stellar Tracks and Isochrones'' \citep[BaSTI-IAC,][]{Hidalgo2018,Pietrinferni2021} and the ``evolutionary extension to the Spectral Analog of Dwarfs'' \citepalias[SANDee,][]{2024RomanBDs6397} collections. In both cases, we chose the isochrones with [Fe/H] and [$\alpha$/Fe] closest to the spectroscopic values in \citet{M4_spectroscopic}: [Fe/H]$=-1.07$ and [$\alpha$/Fe]$=0.39$. We selected the isochrone with [Fe/H]$=-1.1$ and [$\alpha$/Fe]$=0.35$ from the SANDee grid. We used the BaSTI web interpolator\footnote{\href{http://basti-iac.oa-abruzzo.inaf.it/isocs.html}{http://basti-iac.oa-abruzzo.inaf.it/isocs.html}} to obtain an isochrone with [$\alpha$/Fe]$=0.4$ and the metal mass fraction, $Z$, set to the same value as that in the chosen SANDee isochrone ($Z=0.002237$). In both cases, we used the age of 12 Gyr from \citet{2009BedinM4end} and the distance of 1.85 kpc from \citet{2021BaumgardtGaiaDist}. The optical reddening, $E(B-V)$, was treated as a free parameter with temperature-dependent reddening corrections evaluated using the BasicATLAS \citep{BasicATLAS} software package and $R_V=3.67$ \citep{2012HendricksM4red}. When calculating synthetic photometry, we adopted the net throughput curves of NIRISS\footnote{See the dedicated \href{https://jwst-docs.stsci.edu/jwst-near-infrared-imager-and-slitless-spectrograph/niriss-instrumentation/niriss-filters}{JDox page}.}, which include both the transmissivity of the filter and the optical train of the instrument. For ACS/WFC, we employed the instrument response curves from \texttt{stsynphot}\footnote{\href{https://stsynphot.readthedocs.io/en/latest/}{https://stsynphot.readthedocs.io/en/latest/}}.

As SANDee isochrones are only available at $T_\mathrm{eff}\leq4000\ \mathrm{K}$, we extended them to cover the entire range of available photometry by calculating additional model atmospheres using ATLAS-9 \citep{ATLAS9_1,ATLAS9_2,SYNTHE} and evolutionary models using MESA \citep{MESA,MESA_2,MESA_3,MESA_4,MESA_5}, following the procedure described in \citet{roman_47tuc}. The two isochrones evaluated at the best-fit reddening are plotted in Figure~\ref{fig:m4iso1}. The best-fit reddening for each isochrone in each filter combination is indicated in the figure legend. Both isochrones suffer from two shortcomings:
\begin{enumerate}
    \item The best-fit reddening, $E(B-V)$, exceeds the values in the previous studies of bright cluster members (e.g., $E(B-V)=0.37$ in \citet{2012HendricksM4red}, $E(B-V)=0.35$ in \citet{1996AJ....112.1487H}), by $\gtrsim 0.1$ mag. It must be noted that $E(B-V)$ varies by up to $0.1$ mag depending on the specific line of sight within the cluster \citep{2012HendricksM4red}, and the large best-fit $E(B-V)$ value derived in this work may be partly explained by above-average reddening in the observed field. Moreover, the NASA/IPAC Galactic Reddening service\footnote{\href{https://irsa.ipac.caltech.edu/applications/DUST/}{https://irsa.ipac.caltech.edu/applications/DUST/}} suggests that $E(B-V)$ may be as high as $0.43$ in the observed part of the cluster. However, the inferred $E(B-V)$ of $\sim 0.5$ from the $(m_{\rm F814W} - m_{\rm F150W})$ CMD (middle panel of Figure~\ref{fig:m4iso1}) is still irreconcilable with optical estimates in the literature even for the most extreme lines of sight.
    \item In all CMDs, at faint magnitudes ($m_{\rm F150W}$ $\gtrsim 20$) the observed color--magnitude trend appears redder than the colors predicted by the best-fit isochrones.
\end{enumerate}

\begin{figure}
    \centering
    \includegraphics[width=0.9\columnwidth]{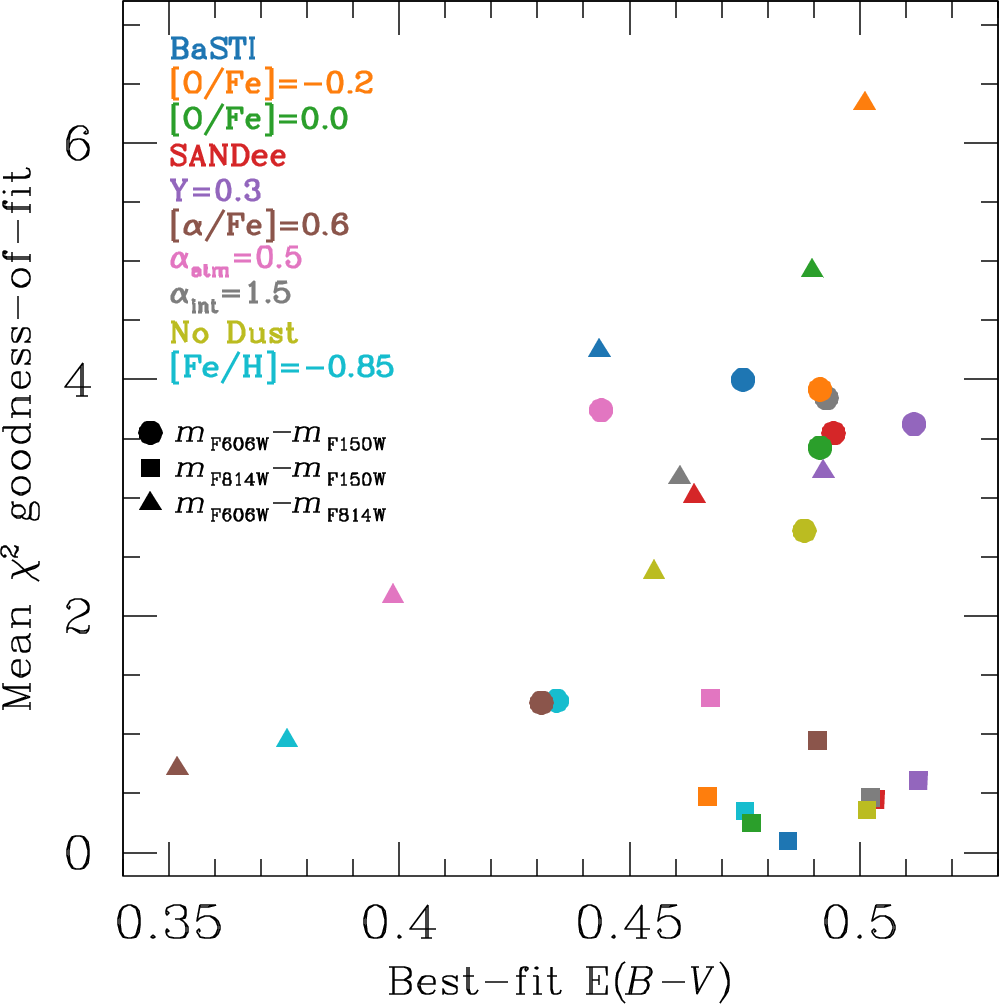}
    \caption{Goodness-of-fit parameter ($\chi^2$) and the best-fit E($B-V$) for each isochrone tested in our work (see the text for details). The most accurate models are expected to have low $\chi^2$, and best-fit $E(B-V)$ that are both comparable to the values inferred from bright members in the literature ($0.3-0.4$) and consistent across the three considered color combinations.}
    \label{fig:m4iso2}
\end{figure}

To determine the origin of these discrepancies, we investigated a wide variety of possible causes by recalculating the SANDee isochrone with alternative parameters. For all test isochrones, the goodness-of-fit value and the best-fit reddening are plotted in Figure~\ref{fig:m4iso2}. The test isochrones that differ the most from the nominal model are overplotted on the observed photometry in Figure~\ref{fig:m4iso3}. The following possible explanations for the observed discrepancy were explored:
\begin{enumerate}
    \item Reduced oxygen abundance. Globular clusters are characterized by the oxygen-sodium anti-correlation \citep{NaO_anticorrelation}, with oxygen-deficient stars typically comprising the majority of members \citep{population_ratios}. At low effective temperatures, [O/Fe] controls the abundance of water vapor that produces prominent absorption bands in the infrared. We calculated two test isochrones with [O/Fe]$=0.0$ and [O/Fe]$=-0.2$ (we note that the nominal abundance of oxygen is [O/Fe]=[$\alpha$/Fe]$=$0.35). Both isochrones display more aggressive reddening of the lower MS in $(m_{\rm F814W} - m_{\rm F150W})$, as expected. However, the fit is much worse in $(m_{\rm F606W} - m_{\rm F814W})$, as these bands do not overlap with major $\mathrm{H}_2 \mathrm{O}$ absorption bands, and the reduced average opacity of the atmosphere increases its effective temperature, making the optical color bluer. The oxygen abundance therefore cannot explain the observed discrepancy on its own, but it may be able to do so in conjunction with another effect that compensates for the decrease in the average opacity.
    \item Increased $\alpha$-enhancement or metallicity. We compared our CMDs to the SANDee isochrone with [Fe/H]$=-0.85$ ($0.25$ dex more metal-rich than the nominal case) and computed a test isochrone with [$\alpha$/Fe]$=0.6$ ($0.25$ dex more $\alpha$-enhanced than the nominal case). The effects of both metallicity and $\alpha$-enhancement on the observed CMD are similar: they increase the average atmospheric opacity, resulting in a reduction in the effective temperature and redder colors. The only exception is the effect of [$\alpha$/Fe] in $(m_{\rm F814W} - m_{\rm F150W})$, where the influence of absorption bands within the filter wavelengths dominates over the effect of average opacity, resulting in a slightly bluer color than the nominal SANDee isochrone. Both test isochrones yield a better fit to the data and a lower value of $E(B-V)$ in accordance with the literature. If the increased metallicity and/or $\alpha$-enhancement is indeed responsible for the observed reddening of the lower MS, the atmospheric composition must be varying with stellar mass in order to remain consistent with the spectroscopic chemistry inferred from observations of higher-mass members. We also note that increasing [Fe/H] or [$\alpha$/Fe] appears to increase the scatter in the best-fit $E(B-V)$ values among our CMDs (see Figure~\ref{fig:m4iso2}), suggesting that [Fe/H] and [$\alpha$/Fe] adjustments are not sufficient to explain the discrepancy on their own.

\begin{figure}
    \centering
    \includegraphics[width=\columnwidth]{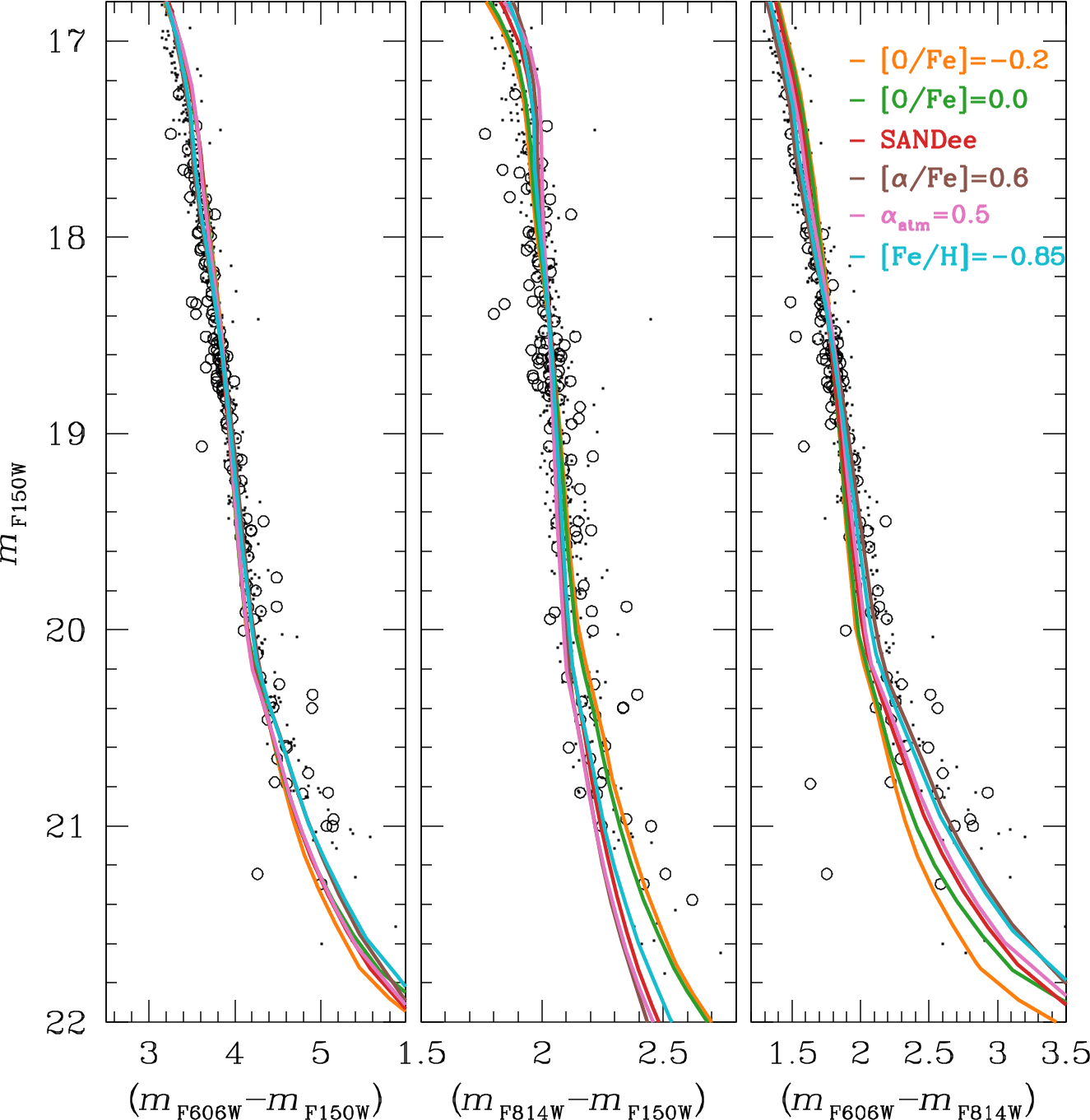}
    \caption{Selected alternative isochrone fits that are most different from the nominal SANDee isochrone shown in Fig.~\ref{fig:m4iso1}. The nominal isochrone is shown in gray for reference. Depending on the filter combination, some SANDee isochrones provide a good fit to the observed CMD, even at the faint-end of the MS, but no isochrone can capture all features in all three CMDs.}
    \label{fig:m4iso3}
\end{figure}

    \item Increased helium mass fraction. The nominal SANDee isochrone adopts the helium mass fraction of $Y=0.25$. We calculated a test isochrone with $Y=0.3$ to evaluate the effect of helium content on stellar evolution. The effect of the helium mass fraction on the lower main sequence is comparatively small and mostly degenerate with the effect of stellar mass. It therefore cannot be observed in the CMD. On the upper main sequence and near the turn-off point, higher $Y$ results in bluer colors and a correspondingly higher best-fit value of $E(B-V)$, exacerbating the discrepancy.
    \item Convective mixing length. In SANDee model isochrones, convective transfer in both stellar interiors and atmospheres is treated in the formalism of the mixing-length theory \citep{MLT}. In stellar interiors, the mixing length of $\alpha_\mathrm{int}=1.82$ scale heights is adopted, which is based on the solar calibration in \citet{solar_alpha}. In stellar atmospheres, SANDee employs a mixing length formula from \citet{ML_formula} derived from radiation hydrodynamics simulations. In the ATLAS models calculated in this study to extend SANDee to $T_\mathrm{eff}>4000\ \mathrm{K}$, we kept the default ATLAS atmospheric mixing length of $\alpha_\mathrm{atm}=1.25$ scale heights, which is inferred from the solar calibration \citep{ATLAS_alpha}. To investigate the effect of convection on the isochrone, we computed two test isochrones with $\alpha_\mathrm{int}=1.5$ and $\alpha_\mathrm{atm}=0.5$. Changing the interior mixing length does not noticeably alter the isochrone below the MS turn-off, and cannot explain the observed discrepancies. Reducing the atmospheric mixing length impacts the atmospheric structure in a similar way to increased metallicity or $\alpha$-enhancement; however, the magnitudes of these effects are only comparable near $T_\mathrm{eff}\sim4000\ \mathrm{K}$. At lower temperatures, where we see the largest discrepancy between the data and the model, the effect of convective mixing length is much smaller and cannot explain our observations.
    \item Dust condensation. In the atmospheres of cool stars and brown dwarfs, liquid and solid species may condense out of the gaseous form, offsetting the chemical equilibrium and altering the spectral energy distribution. In SANDee, at the range of $T_\mathrm{eff}$ considered in this study, dust condensation is treated using the inefficient settling (equilibrium) model from \citet{PHOENIX_dust}. We calculated an additional test isochrone without dust condensation. While the dust-free isochrone predicts redder colors at the lowest effective temperatures, the effect of condensation at the low metallicity of the cluster and the comparatively warm temperatures of stars within our photometric limit ($T_\mathrm{eff}>3000\ \mathrm{K}$) is not sufficient to account for the observed discrepancy.
\end{enumerate}
The tests described above suggest that it may be possible to reconcile the observed CMDs and theoretical models by simultaneously reducing the oxygen abundance by $\sim 0.4\ \mathrm{dex}$ in low-mass stars to capture the unexpectedly large reddening of the $(m_{\rm F814W} - m_{\rm F150W})$ color near the end of the MS, and increasing the average opacity of the atmosphere to reduce the effective temperatures of low-mass stars. The latter effect is most readily attained by increasing [Fe/H] and/or [$\alpha$/Fe] by $\sim 0.3\ \mathrm{dex}$, and may be indicative of incomplete opacities in the model atmospheres. However, this seems unlikely, as the effect is consistent across multiple modeling codes, and persists into relatively high temperatures of $T_\mathrm{eff}\gtrsim 4000\ \mathrm{K}$, where the opacity model is more straightforward. Other effects such as dust condensation, helium enhancement and convection cannot explain the observed discrepancy.

If not in the isochrones, the problem may reside in the data, although we do not find any clear systematic issue that can explain the discrepancies. Possible speculations include color-dependent systematics or a second-order by-product of the brighter-fatter effect. For the latter case, the NIRISS ePSFs are made using bright stars that might partially be affected by the brighter-fatter effect \citep[although specific magnitude thresholds were adopted in the ePSF modeling to avoid this issue;][]{2023LibralatoNIRISS}. The faintest stars in our CMDs, which instead are not impacted by the brighter-fatter issue, could result in small magnitude-dependent systematics when fitted with these ePSFs, given that the ePSFs are not the real representation of the flux distribution of such faint sources. The main flaw in this hypothesis is the good agreement between the photometry of the Northern and Southern fields, which were taken with different readout patterns and exposure times and hence are impacted differently by the brighter-fatter phenomenon.

From the physical side, a possible explanation could be that, at the faint end of our CMDs, the points are not redder than the canonical theoretical tracks but brighter, and these brighter sources are just MS binaries made by very-low-mass MS stars. However, it is unclear as to why such objects should comprise the majority of the sources at these magnitudes and which dynamical processes could preferably retain them at the clustercentric distance of our fields. A future analysis of the NIRCam data of NGC~6121 will allow us to shed more shed light on the topic.

\begin{figure}
    \centering
    \includegraphics[width=0.75\columnwidth]{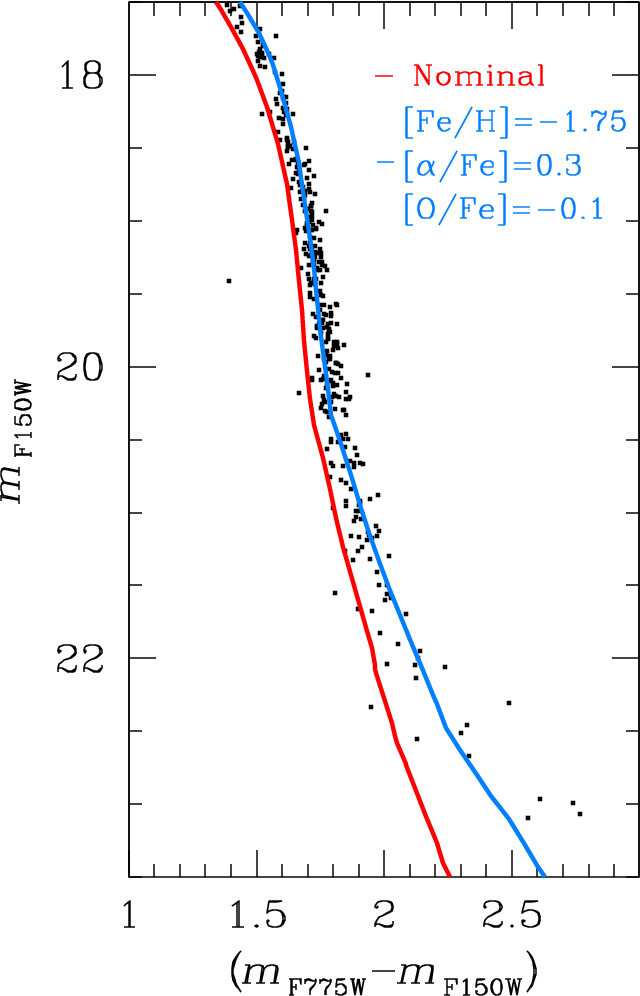}
    \caption{CMD for the MS of NGC~6397. The red line represents an isochrone with nominal parameters from the literature, whereas the blue line shows the adjusted option that provides the best fit to the data (see the text for details).}
    \label{fig:ngc6397iso}
\end{figure}

\subsection{NGC~6397}

For NGC~6397, we followed the procedure in \citet{roman_47tuc} to calculate two model isochrones: a nominal isochrone with parameters matching the literature, and a corrected isochrone with adjusted [Fe/H], [$\alpha$/Fe] and [O/Fe] to attain the best correspondence between the model and the data. For the nominal isochrone, we adopted the median abundances of elements from \citet{2012ApJ...745...27M} ([C/Fe]$=0.03$, [N/Fe]$=0.58$, [O/Fe]$=0.28$), age of 13.5 Gyr from beryllium dating in \citet{beryllium_dating}, metallicity of [Fe/H]$=-1.88$, distance modulus of $12.02$, and reddening of $E(B-V)=0.22$ from \citet{2018ApJ...864..147C}, and $R_V=3.1$. As seen in Fig.~\ref{fig:ngc6397iso}, the nominal isochrone suffers from the same shortcomings as the nominal isochrones of NGC~6121: it appears consistently bluer than the data across the entire observed mass range, and fails to reproduce the rapid reddening of the CMD at $m_{\rm F150W}$ $\gtrsim 21$.

We attempted to correct the nominal isochrone by introducing similar offsets to its parameters, as inferred in the case of NGC~6121. Namely, we reduced [O/Fe] by $\sim 0.4$ dex to [O/Fe]$=-0.1$ and increased [$\alpha$/Fe] by $0.3$ dex. We then increased the metallicity to [Fe/H]$=-1.75$ to resolve the remaining discrepancy in color. This corrected isochrone provides a better fit to the data. However, similarly to the case of NGC~6121, we find that the corrected isochrone does not fully reproduce the red turn at the bottom of the main sequence. As discussed in the previous section, we do not know whether this is a real feature --the ingredients of which are not properly included in the current theoretical tracks-- or a product of a systematic in the data.

\subsection{The WD cooling sequence}

\begin{figure*}
    \centering
    \includegraphics[height=6.85cm, keepaspectratio]{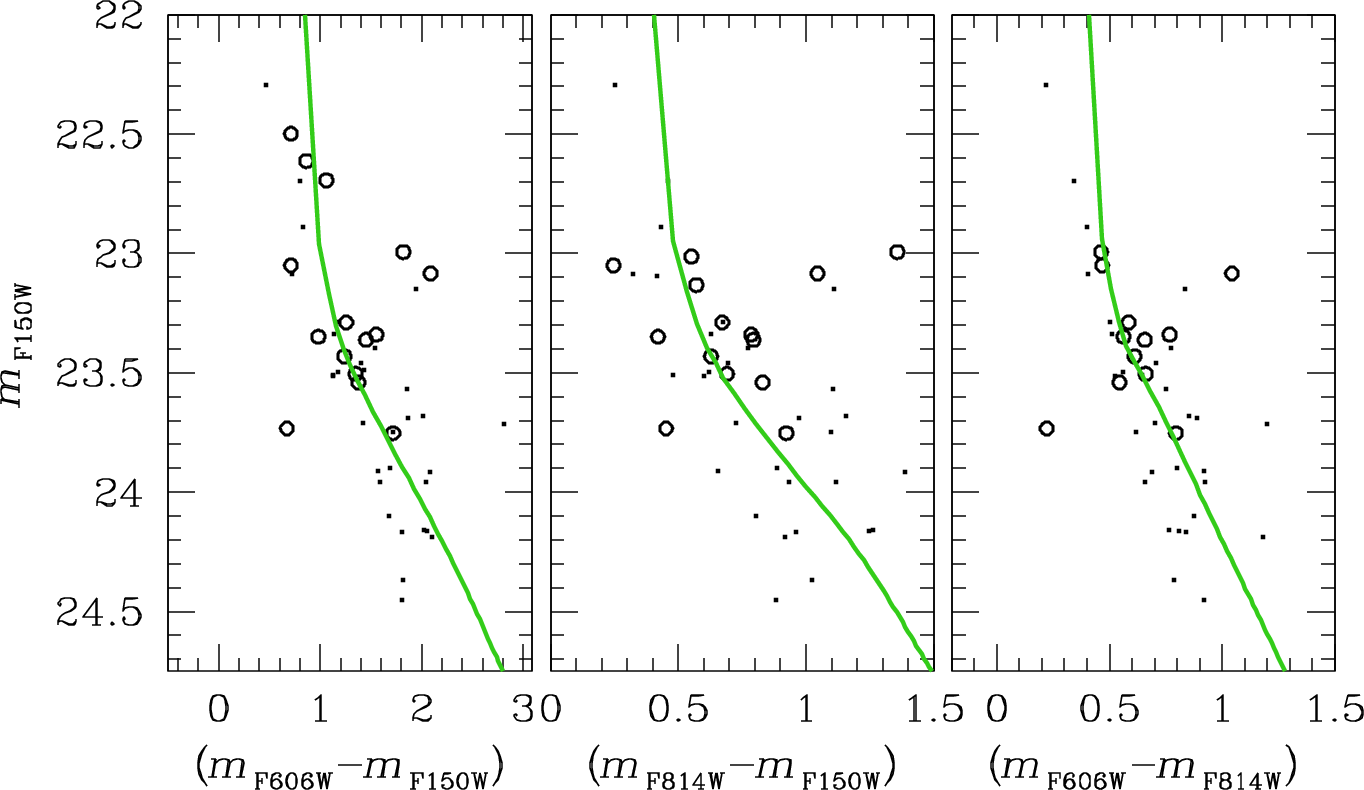}
    \hspace{0.5 cm}
    \includegraphics[height=6.73cm, keepaspectratio]{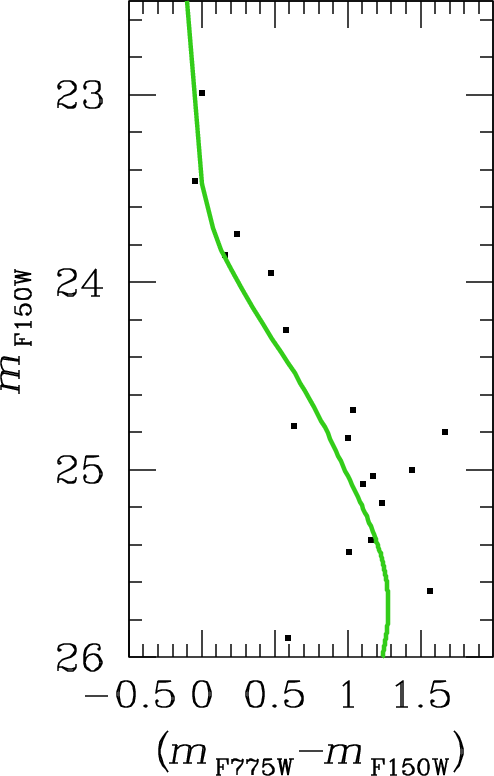}
    \caption{Collection of CMDs focused on the WD sequence of NGC~6121 (three leftmost panels) and NGC~6397 (rightmost panel). Only well-measured cluster members are shown. For NGC~6121, we use the same symbols as in the previous figures. A BaSTI WD isochrone is shown in each plot (green line).}
    \label{fig:wdiso}
\end{figure*}

Figure~\ref{fig:wdiso} zooms onto the WD region to show a comparison with a set of BaSTI WD isochrones \citep{2010SalarisWD}. The reddening, age, and distance modulus are the same as in Figs.~\ref{fig:m4iso1} and \ref{fig:ngc6397iso}. The limited depth of the \hst data (see Appendix~\ref{sec:artstar}) makes it possible to study only the bright part of the WD sequence, that is, WDs with a mass of $\sim$0.55 $M_\odot$. Nevertheless, there is a good agreement between the observed sequence and the theoretical track.

\section{Luminosity and mass functions}\label{sec:lfmf}

Thanks to the PM-cleaned CMDs made by combining \hst and \jwst data, we studied the MS of the clusters down to $<$0.1 $M_\odot$ and computed the present-day local LFs and MFs of the two clusters. The LF and MF of a stellar population provide useful insights about the formation and dynamical evolution of the system \citep[e.g.,][]{1997VesperiniIMF}.

\begin{figure*}
    \centering
    \includegraphics[width=0.975\columnwidth]{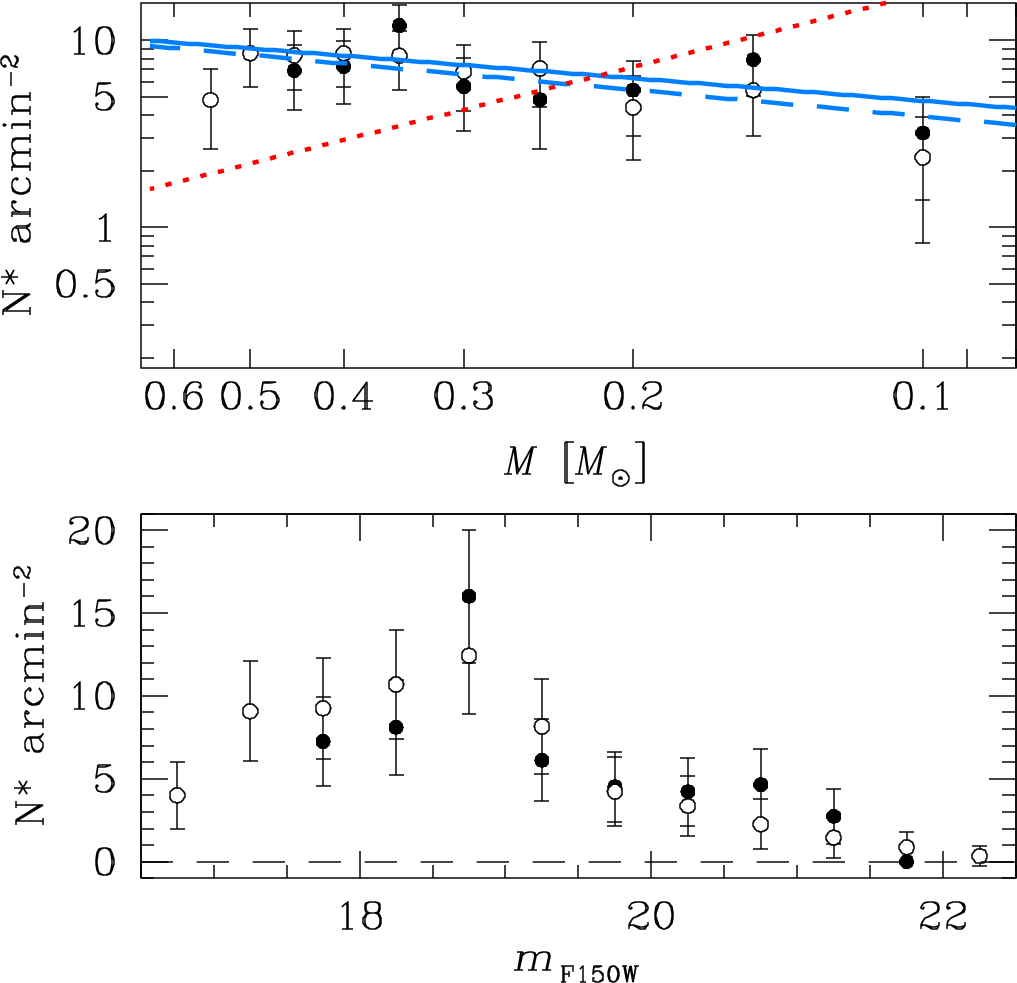}
    \hspace{0.5 cm}
    \includegraphics[width=0.975\columnwidth]{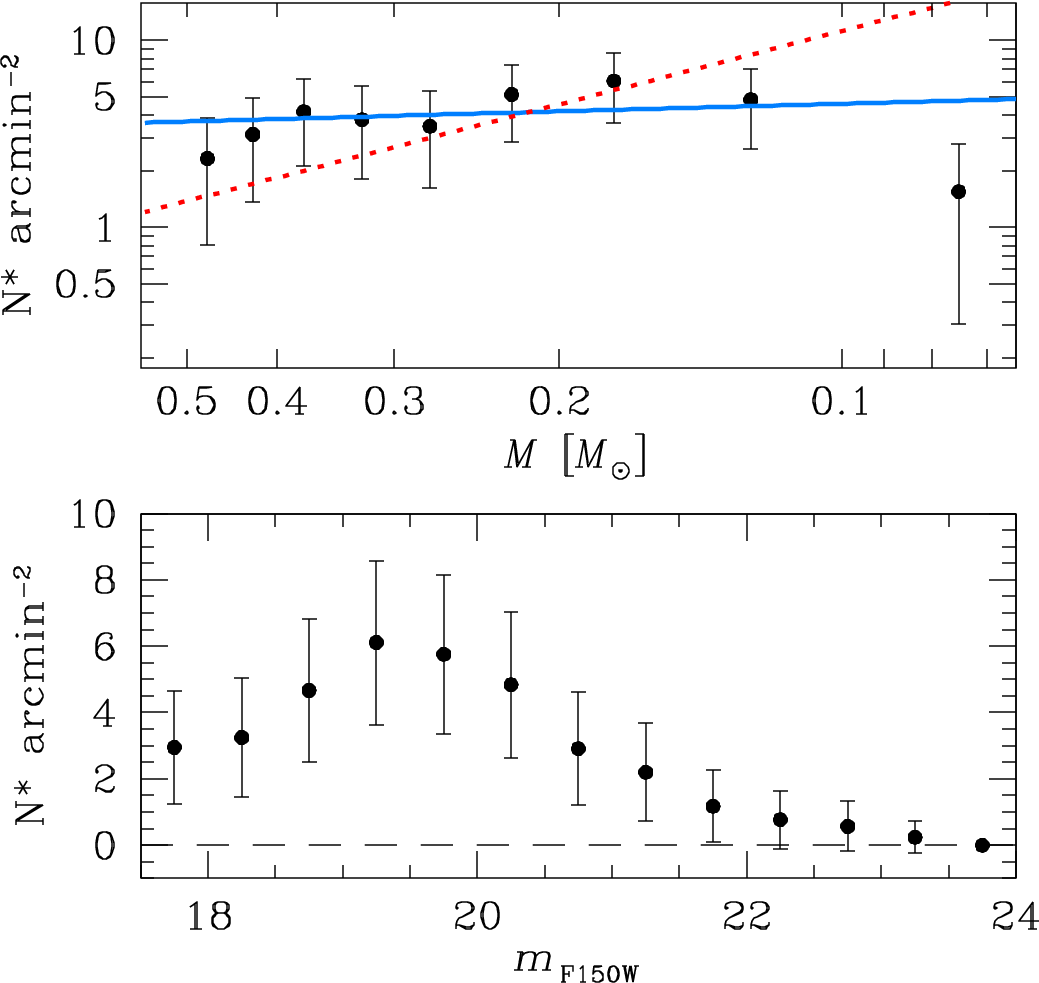}
    \caption{Present-day local LFs (bottom panels) and MFs (top panels) for NGC~6121 (left column) and NGC~6397 (right column). Error bars represent the propagated Poisson errors. The horizontal, dashed line in the bottom panels is set to zero for reference. For NGC~6121, we plot the results for the Northern and Southern fields with open and filled dots, respectively. In the top panels, the blue lines represent straight-line fits to the points, the slopes of which are reported in the text. For the case of NGC~6121, the blue solid line refers to the Northern field, while the dashed line is obtained from the Southern-field points. Finally, the red, dotted lines have a slope of $-1.3$ (i.e., that of a Kroupa MF in the observed mass range) and arbitrary intercept to fit in the plot.}
    \label{fig:lfmf}
\end{figure*}

The LF of each cluster was obtained as follows. We selected a sample of well-measured (see Sect.~\ref{sec:iso}) MS stars, divided it into bins of 1 mag in width (with steps of 0.5 mag) and counted the number of sources in each bin. The number of stars per bin was corrected for completeness using the outputs of the artificial-star tests described in Appendix~\ref{sec:artstar}. Finally, we normalized the number of stars by the total area in arcmin$^2$ of the overlapping \hst and \jwst fields. This way, we can directly compare the LFs of the two clusters, and also cross-checked the photometry of the Northern and Southern fields of NGC~6121 again, which were analyzed independently. The present-day local LFs of NGC~6121 and NGC~6397 are shown in the bottom panels of Fig.~\ref{fig:lfmf}. The field contamination can be a source of uncertainty when computing the LF of a cluster. However, these two systems are the closest GCs to the Sun, and the separation between cluster and field objects in the VPD was found to be clear even at the faintest magnitudes; we neglected the contribution of the field contamination.

Thanks to the isochrones described in Sect.~\ref{sec:iso}, we were able to compute the present-day local MFs in a similar fashion. We converted the F150W-filter magnitudes in masses using the isochrones and counted the number of stars in bins of 0.1 $M_\odot$ in width with steps of 0.05 $M_\odot$. As for the LFs, the number of stars was corrected for completeness and normalized by the area of the field. The result is shown in the top panels of Fig.~\ref{fig:lfmf}. Both clusters show a change of slope in the MFs. For NGC~6121, the break point happens at $\sim$0.3--0.4 $M_\odot$. For NGC~6397, the slope change is slightly below 0.2 $M_\odot$. We report the values of the LFs and MFs for the two clusters in Tables~\ref{tab:m4lfmf} and \ref{tab:ngc6397lfmf}.

We measured the slope $\alpha$ of the present-day local MFs\footnote{We consider the expression $N(m) \propto m^{\alpha}$, i.e., $\alpha = -2.35$ for a Salpeter MF.} by fitting a straight line to the all mass bins in Fig.~\ref{fig:lfmf} and found:
\begin{itemize}
    \item NGC~6121 (0.1$<$$M$$<$0.6 $M_\odot$): $\alpha_{\rm @ 3.6\,arcmin} = 0.40 \pm 0.29$;
    \item NGC~6121 (0.1$<$$M$$<$0.6 $M_\odot$): $\alpha_{\rm @ 4.8\,arcmin} = 0.47 \pm 0.18$;
    \item NGC~6397 (0.07$<$$M$$<$0.5 $M_\odot$): $\alpha_{\rm @ 11.5\,arcmin} = -0.14 \pm 0.26$.
\end{itemize}
For both clusters, the measured slopes are significantly flatter than the values expected in the observed mass range for a \citet[$\alpha^{\rm Salpeter} = -2.35$]{1955SalpeterMF} or a \citet[$\alpha^{\rm Kroupa} = -1.3$]{2001KroupaIMF} stellar MF. A flattening of the global stellar MF is the expected consequence of the preferential loss of low-mass stars due to the effects of two-body relaxation \citep[see, e.g.,][]{1997VesperiniIMF}. For the interpretation of the observed values of the slope of the MF, however, it is important to consider that effects other than the stellar escape can alter the global slope of the MF. For example, two-body relaxation causes massive stars to segregate in the cluster's inner regions and low-mass stars to migrate outwards, leading to a variation of the MF slope with the distance from the cluster center. As massive stars migrate inwards and low-mass stars outwards, the slope of the MF in the inner (outer) regions becomes flatter (steeper) than the global one \citep{1997VesperiniIMF}. We refer to, for example, \citet{1995KingGC} for an early \hst study of the radial variation of the MF in NGC~6397, or the recent study by \citet[and its online database]{2023BaumgardtMF} for the radial variation of the MF in GCs.

The values of $\alpha$ found in NGC~6121 and NGC~6397 are the result of all these processes affecting both the global and the local MF. The location of the fields for NGC~6121 and that of NGC~6397 correspond, respectively, to an intermediate region of the clusters for NGC~6121 (about 0.8 $r_{\rm h}$ and 1 $r_{\rm h}$ for Northern and Southern fields, respectively) and an outer region for NGC~6397 (at about 4$r_{\rm h}$ from the cluster's center). Part of the difference between the observed slope of the MF in NGC~6121 and NGC~6397 is therefore due to the different regions observed for the two clusters and is consistent with the expected trend between the slope and the distance from the cluster center. Part of the difference may also be due to differences between the global MFs of the two clusters. For both clusters, the flat values of the slope of the MF revealed by our observations suggest that these systems have been significantly affected by mass loss and the preferential escape of low-mass stars. This conclusion is in agreement with the results of a number of previous studies wherein specific N-body and Monte Carlo evolutionary models were built for these two clusters \citep{2008HeggieM4,2008Hurley6397,2009Giersz6397}.

\section{Discussion and conclusions}\label{sec:conc}

The \jwst is ideal for studying low-mass stars because they are faint and red objects, and are brighter at longer wavelengths than their more-massive counterparts, which makes them ideal targets for near-infrared observations. Since the launch of \jwst, an increasing number of authors have begun to focus on faint stars in GCs. Our paper is the fourth in a series investigating the faintest members of NGC~6121 and NGC~6397, the two geometrically closest GCs to us. In this study, we analyzed the parallel fields observed with the NIRISS detector of \jwst.

We discuss the data reduction of the NIRISS data, from the pipeline setup to the extraction of accurate and precise positions and fluxes for all sources in the field. We demonstrate that, in combination with archival \hst data, it is possible to clearly disentangle cluster members from field stars down to $<$0.1 $M_\odot$. The NIRISS data alone have the potential to probe even fainter stars along the MS but the \hst data limit the complete exploitation of the information provided by \jwst. Future follow ups with \jwst will allow us to potentially investigate stars down to the hydrogen-burning limit, as done for the same clusters with the NIRCam data in the primary fields of this project.

We compare the observed CMDs with theoretical isochrones. While there is a good agreement on the brighter part of the MS, the faint-end of the MS is not reproduced by the theoretical tracks using literature values for the extinction and chemical composition. Only models with higher reddening and/or metallicity are able to fit the red turn of the MS. We discuss various hypotheses that might explain these findings, from shortcomings in the isochrones to systematics in the data.

Finally, we compute the LFs and MFs for the two clusters. The slopes of the MF are measured close to the $r_{\rm h}$ for NGC~6121 and in the outer regions at about 4$r_{\rm h}$ for NGC~6397. The differences between the slopes found for the two clusters are generally consistent with the expected effects associated with the spatial outward (inward) migration of low-mass (high-mass) stars due to two-body relaxation (differences between the global MFs of the two clusters could also be in part responsible for the observed differences between the local MFs). In general, our observations reveal a significant flattening in the MF that indicates that, for both of these clusters, the present-day MF has been significantly affected by the preferential loss of low-mass stars.

\begin{table}[t!]
    \centering
    \caption{LF and MF for NGC~6121.}
    \begin{tabular}{c|cc|cc}
        \hline
        \hline
        & \multicolumn{2}{c}{Northern field} & \multicolumn{2}{c}{Southern field} \\
        \hline
        $m_{\rm F150W}$ & N$^\ast$\,arcmin$^{-2}$ & Error & N$^\ast$\,arcmin$^{-2}$ & Error \\
        \hline
        16.75 &       &      &  4.00 & 2.00 \\
        17.25 &       &      &  9.08 & 3.01 \\
        17.75 &  7.25 & 2.69 &  9.26 & 3.04 \\
        18.25 &  8.10 & 2.85 & 10.69 & 3.27 \\
        18.75 & 16.02 & 4.00 & 12.42 & 3.52 \\
        19.25 &  6.12 & 2.47 &  8.16 & 2.86 \\
        19.75 &  4.52 & 2.13 &  4.24 & 2.06 \\
        20.25 &  4.22 & 2.05 &  3.36 & 1.83 \\
        20.75 &  4.64 & 2.15 &  2.26 & 1.50 \\
        21.25 &  2.75 & 1.66 &  1.45 & 1.21 \\
        21.75 &       &      &  0.88 & 0.94 \\
        22.25 &       &      &  0.34 & 0.59 \\
        \hline
        \multicolumn{5}{c}{ }\\
        \hline
        \hline
        $M$\,[$M_\odot$] & N$^\ast$\,arcmin$^{-2}$ & Error & N$^\ast$\,arcmin$^{-2}$ & Error \\
        \hline
        0.10 &  3.20 & 1.79 & 2.37 & 1.54 \\
        0.15 &  7.88 & 2.81 & 5.42 & 2.33 \\
        0.20 &  5.43 & 2.33 & 4.38 & 2.09 \\
        0.25 &  4.81 & 2.19 & 7.09 & 2.66 \\
        0.30 &  5.68 & 2.38 & 6.81 & 2.61 \\
        0.35 & 12.01 & 3.47 & 8.29 & 2.88 \\
        0.40 &  7.26 & 2.69 & 8.54 & 2.92 \\
        0.45 &  6.88 & 2.62 & 8.32 & 2.88 \\
        0.50 &       &      & 8.54 & 2.92 \\
        0.55 &       &      & 4.81 & 2.19 \\
        \hline 
    \end{tabular}
    \label{tab:m4lfmf}
\end{table}      

\begin{table}[t!]
    \centering
    \caption{LF and MF for NGC~6397.}
    \begin{tabular}{c|cc}
        \hline
        \hline
        $m_{\rm F150W}$ & N$^\ast$\,arcmin$^{-2}$ & Error \\
        \hline
        17.75 & 2.95 & 1.72 \\
        18.25 & 3.24 & 1.80 \\
        18.75 & 4.67 & 2.16 \\
        19.25 & 6.11 & 2.47 \\
        19.75 & 5.76 & 2.40 \\
        20.25 & 4.83 & 2.20 \\
        20.75 & 2.91 & 1.71 \\
        21.25 & 2.20 & 1.48 \\
        21.75 & 1.18 & 1.08 \\
        22.25 & 0.77 & 0.87 \\
        22.75 & 0.57 & 0.75 \\
        23.25 & 0.24 & 0.49 \\
        \hline 
        \multicolumn{3}{c}{ }\\
        \hline
        \hline
        $M$\,[$M_\odot$] & N$^\ast$\,arcmin$^{-2}$ & Error \\
        \hline
        0.075 & 1.55 & 1.25 \\
        0.125 & 4.81 & 2.19 \\
        0.175 & 6.07 & 2.46 \\
        0.225 & 5.14 & 2.27 \\
        0.275 & 3.48 & 1.87 \\
        0.325 & 3.77 & 1.94 \\
        0.375 & 4.16 & 2.04 \\
        0.425 & 3.14 & 1.77 \\
        0.475 & 2.34 & 1.53 \\
        \hline 
    \end{tabular}
    \label{tab:ngc6397lfmf}
\end{table}

\section{Data availability}

As part of this publication, we release the star catalogs (positions, photometry, and membership) and the astrometrized stacked images through the CDS and our website\footnote{\href{https://web.oapd.inaf.it/bedin/files/PAPERs_eMATERIALs/JWST/GO-1979/P04/}{https://web.oapd.inaf.it/bedin/files/\-PAPERs\_eMATERIALs/JWST/GO-1979/P04/}}. A description of this supplementary online electronic material is provided in Appendix~\ref{sec:release}.

\begin{acknowledgements}
ML thanks Paul Goudfrooij for his suggestions about the NIRISS data reduction. ML also thanks Marshall Perrin and Marcio Melendez Hernandez for providing extended WebbPSF PSF models for NIRISS used by \kstwo to make the bright-star masks. This work is based on funding by: INAF under the WFAP project, f.o.:1.05.23.05.05; The UK Science and Technology Facilities Council Consolidated Grant ST/V00087X/1; and STScI NASA funding associated with GO-1979. Based on observations with the NASA/ESA/CSA \textit{JWST}, obtained at the Space Telescope Science Institute, which is operated by AURA, Inc., under NASA contract NAS 5-03127. Also based on observations with the NASA/ESA \textit{HST}, obtained at the Space Telescope Science Institute, which is operated by AURA, Inc., under NASA contract NAS 5-26555. This work has made use of data from the European Space Agency (ESA) mission {\it Gaia} (\url{https://www.cosmos.esa.int/gaia}), processed by the {\it Gaia} Data Processing and Analysis Consortium (DPAC, \url{https://www.cosmos.esa.int/web/gaia/dpac/consortium}). Funding for the DPAC has been provided by national institutions, in particular the institutions participating in the {\it Gaia} Multilateral Agreement. This research made use of \texttt{astropy}, a community-developed core \texttt{python} package for Astronomy \citep{astropy:2013, astropy:2018}.
\end{acknowledgements}

\bibliographystyle{aa}
\bibliography{JWST_GO_1979_NIRISS}

\clearpage
\begin{appendix}

\section{The absolute PM of NGC~6397}\label{appendix:abspm}

The many background galaxies clearly visible in both the \hst and \jwst stacked images of NGC~6397 offer us the opportunity to estimate the absolute PM of the cluster \citep[e.g.,][]{2006MiloneNGC6397,2013MassariM70,2018LibralatowCen}. This simple exercise is meant to show the goodness of our data reduction and another example of the astrometric potential of the \hst-\jwst duo.

We identified by eye a sample of 20 background galaxies with a point-like core in the NIRISS stacked images. These sources are tightly distributed in the VPD (green crosses in Fig.~\ref{fig:ngc6397overview}). The 3$\sigma$-clipped median value of the PM in each coordinate of these sources, with opposite sign, is the absolute PM of the cluster in our NIRISS field:
\begin{displaymath}
  (\mu_\alpha \cos\delta,\mu_\delta)_{\rm @ Field} = (3.309 \pm 0.145,-17.742 \pm 0.132) \phantom{1} \textrm{mas yr}^{-1} \phantom{1} .
\end{displaymath}

Our field is $\sim$11.5 arcmin away from the center of NGC~6397, so the estimate above needs to be corrected by the projection effects that arise from the different line of sight between our analyzed field and the center of mass (COM) of the cluster. Using the equations provided in \citet{2002AJ....124.2639V} and assuming the position of our field and of the cluster's center described in Sect.~\ref{sec:data}, we obtain:
\begin{displaymath}
  (\mu_\alpha \cos\delta,\mu_\delta)_{\rm COM} = (3.257 \pm 0.145,-17.750 \pm 0.132) \phantom{1} \textrm{mas yr}^{-1} \phantom{1} .
\end{displaymath}
The rotation in the plane of the sky of NGC~6397 is small \citep[e.g.,][]{2018BianchiniRot,2021VasilievGCkin}, so we neglected its contribution. Our estimate of the absolute PM of the COM is in agreement with that from the \gaia DR3 provided in the GC database of Holger Baumgardt at the $<$1$\sigma$ level (i.e., $(\mu_\alpha \cos\delta,\mu_\delta)$$=$$(3.251\pm0.005;-17.649\pm0.005)$ \masyr).

\section{Artificial-star tests}\label{sec:artstar}

We computed artificial-star tests to correct for systematic errors
between input magnitude and recovered magnitudes and to estimate the level of completeness in our data. We setup our tests for each cluster as in, for example, \citet{2017BelliniwCenI} and \citet{2023NardiellojwstIII}.

We generated a set of artificial stars with F150W magnitude from close to the saturation limit to $m_{\rm F150W} \sim 33$, i.e., several magnitude below the faintest star in our sample. For the Southern field of NGC~6121, we simulated 100\,000 stars with a colors that lie on a fiducial line representing the MS of the cluster, drew by hand in the observed CMDs with the other F606W and F814W filters. We also simulated 100\,000 stars along the WD sequence and 50\,000 stars with random colors to simulate the distribution of field objects. For the Northern field of NGC~6121, which overlap with existing \hst data is smaller than that between \hst and the Southern field, we reduced the numbers to 40\,000 MS and WD stars and 20\,000 field sources (i.e., about the same number of stars per arcmin$^2$). Finally, for NGC~6397, we instead created 200\,000 stars along the MS, 100\,000 WDs and 50\,000 field objects. All fields analyzed in our work show a rather uniform distribution of objects, thus we assigned to all these stars random positions to uniformly cover the field of view. 

\begin{figure}
    \centering
    \includegraphics[width=\columnwidth]{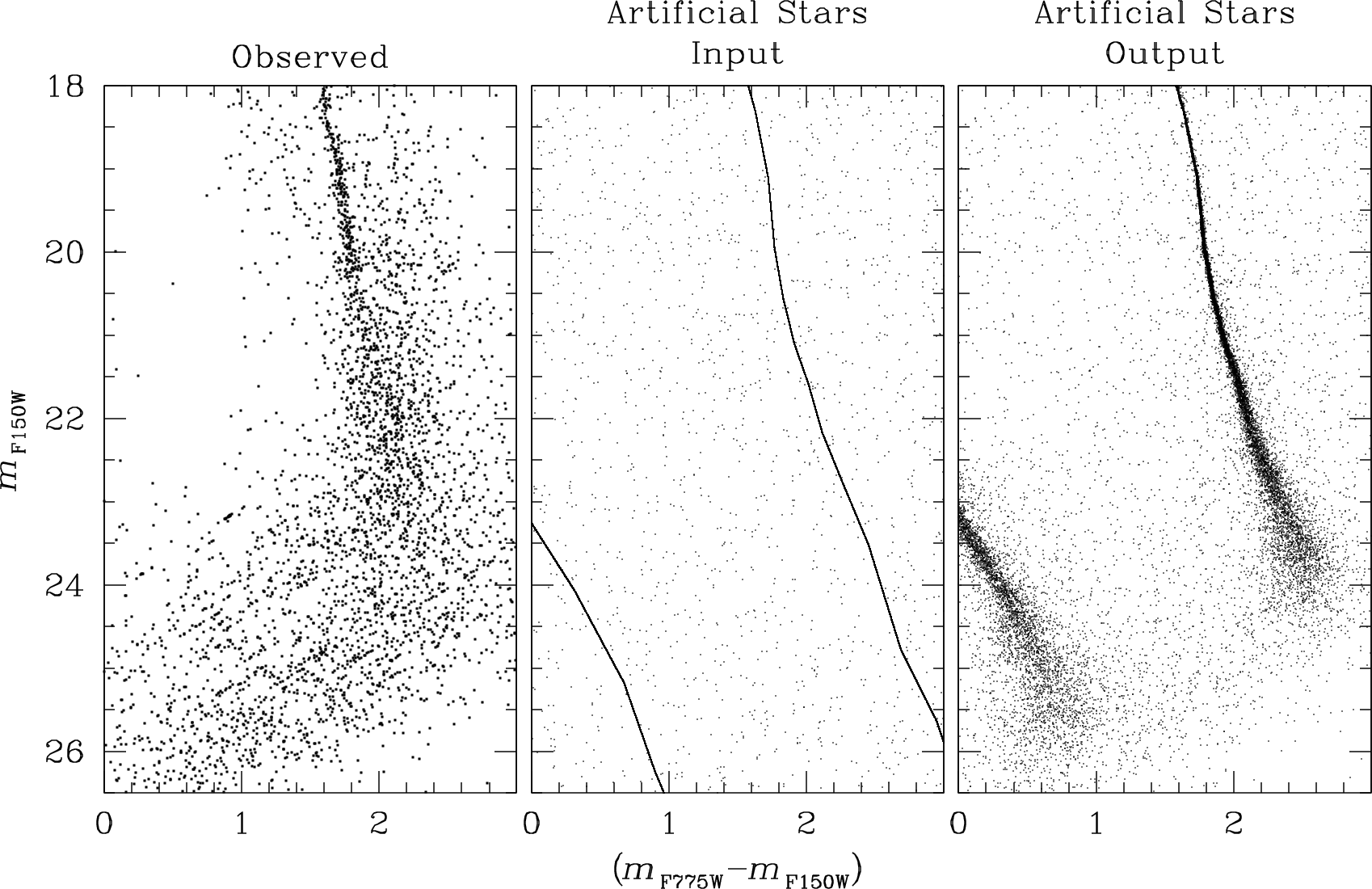}
    \caption{Example of artificial-star test for NGC~6397. (Left): lCMD of the stars measured in the field of NGC~6397. (Middle and Right): CMDs highlighting the artificial stars in input and output from our test, respectively. Only 10\% of the points are shown, for the sake of clarity.}
    \label{fig:ngc6397asinput}
\end{figure}

We then used the \kstwo software to add to each image one artificial star at a time, and then find and measure this source in the same way as done with the real data. Figure~\ref{fig:ngc6397asinput} shows an example for NGC~6397. The CMD with the artificial stars given in input is shown in the middle panel, whereas the CMD made with the \kstwo-based output is shown in the rightmost panel. Both the observed and the output CMDs in this Figure are corrected for the so-called input-output systematic error \citep[e.g.,][]{2019Bedin6752III}, which consists in an increasingly-larger overestimate of the flux of a star towards fainter magnitudes. We corrected the difference between the inserted and recovered values of magnitudes of the artificial stars following the prescriptions of \citet{2009BedinM4end}. In our work, we find that such correction is small ($<$0.1 mag at most) in F150W and it increases for bluer filters, reaching even 0.3 mag for faint stars in the F606W filter. When the completeness drops below 50\%, the input-output correction is less constrained, so for stars fainter than this threshold, the correction was kept fixed at the value of the last reliable estimate. This means that for stars with very low completeness, the correction could be underestimated. All CMDs presented in our work are corrected for this input-output effect.

\begin{figure}
    \centering
    \includegraphics[width=\columnwidth]{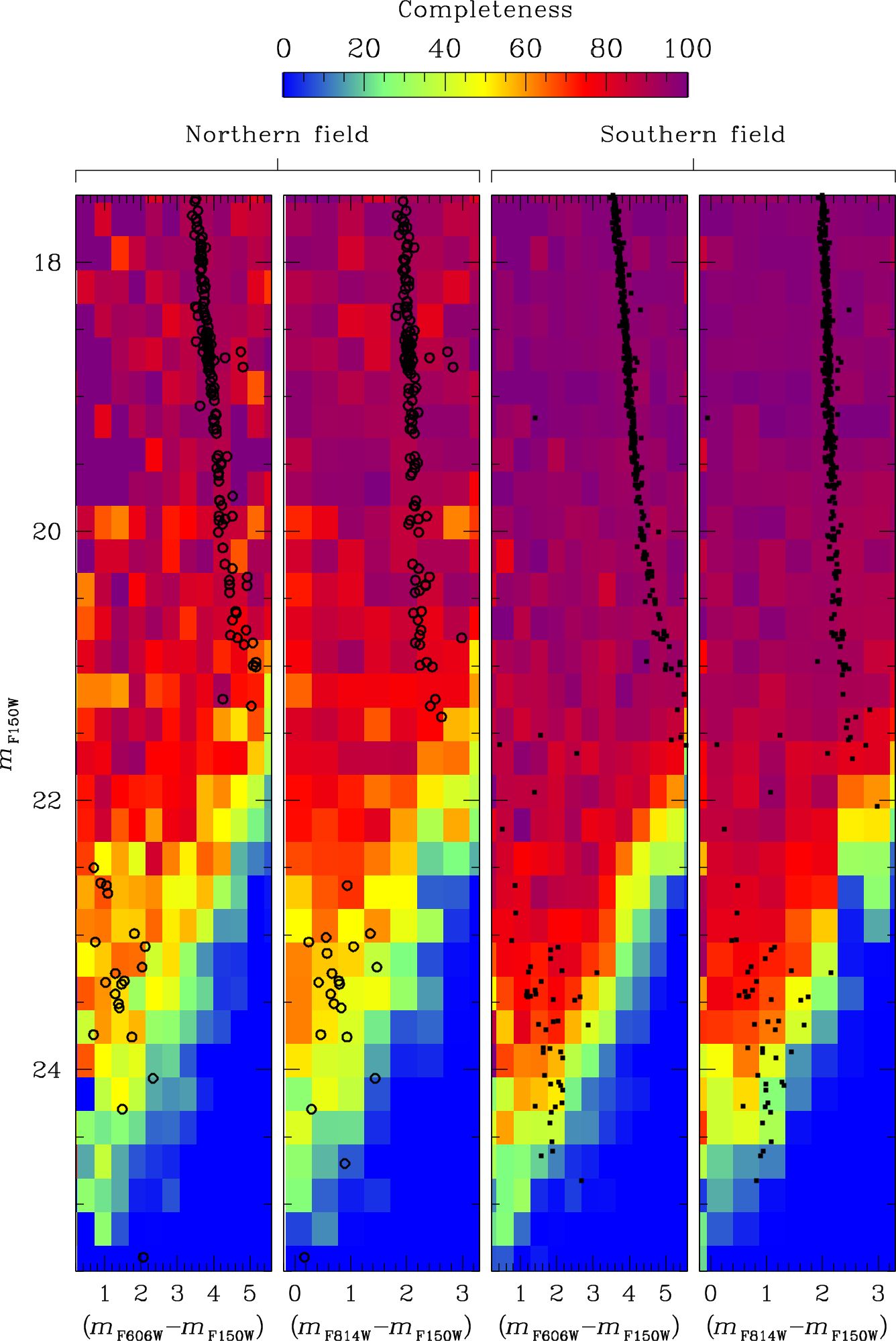}
    \caption{Completeness level for the Northern and Southern fields of NGC~6121. In each CMD, cells are color-coded according to the completeness level of stars in the corresponding magnitude-color interval (see the color bar at the top). Points, with the same symbols as in Fig.~\ref{fig:m4overview}, are the observed cluster members in our data.}
    \label{fig:m4completeness}
\end{figure}

\begin{figure}
    \centering
    \includegraphics[height=12cm, keepaspectratio]{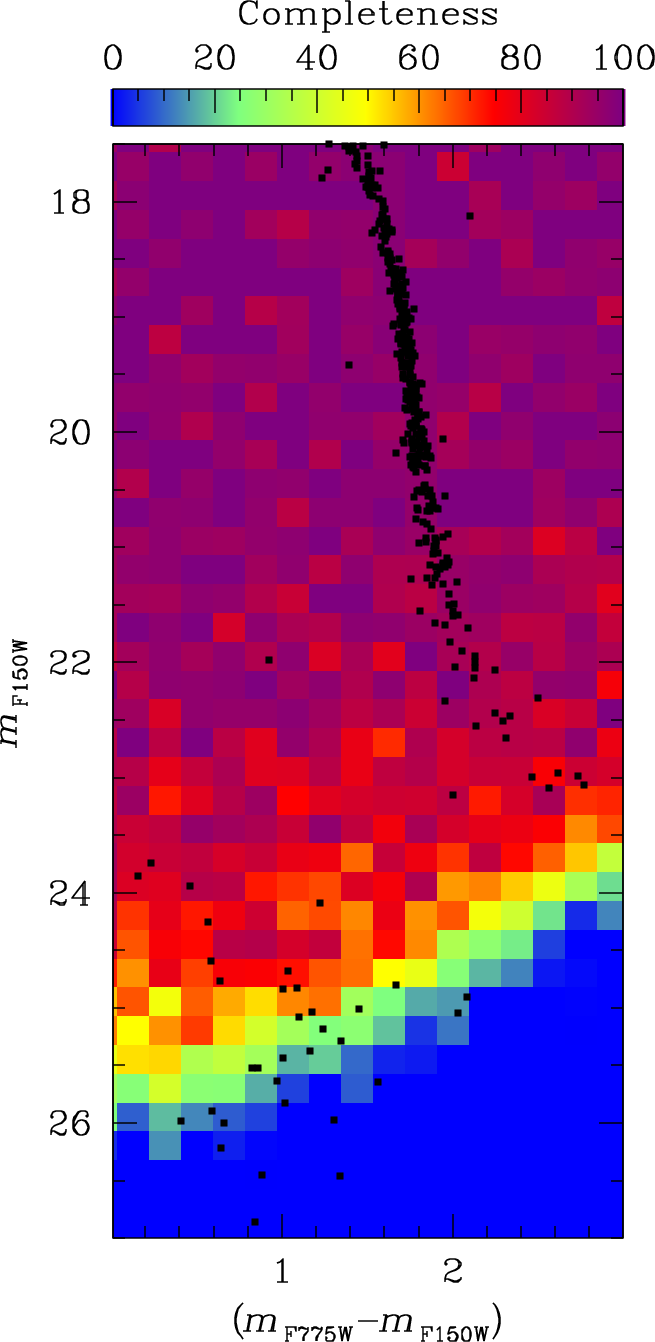}
    \caption{Same as Fig.~\ref{fig:m4completeness} but for NGC~6397.}
    \label{fig:ngc6397completeness}
\end{figure}

The artificial-star tests also allow us to estimate the completeness level in our CMDs. The completeness is defined as the ratio between the number of recovered and injected stars. A star was considered as recovered if the output and input positions differ by less than 0.5 pixel, the difference between output and input magnitudes is less than 0.75 mag, and the star passed the quality selections described in Sect.~\ref{sec:iso}. All these criteria have to be fulfilled in both the \hst and \jwst data. Figures~\ref{fig:m4completeness} and \ref{fig:ngc6397completeness} presents the results for NGC~6121 and NGC~6397, respectively. In both cases, we can see that the number of sources along the MS drops when the completeness is still high, meaning that the density of stars we see in that region of the CMD is real. For NGC~6121, the completeness is higher for the Southern field, where the depth of coverage of both the \hst and \jwst is more uniform. On the other hand, the low completeness along the WD sequence prevents us from studying in detail the WD cooling sequence. Finally, the large cell-to-cell variation in regions of the CMDs not covered by cluster stars, especially in the Northern field of NGC~6121, is due to the relatively-low number of field stars simulated in our test.

The main limitation in our work is the depth of the \hst data. Indeed, if we compute the completeness level of the F150W NIRISS data alone, we find that the 50\% level is at about $m_{\rm F150W} = 25$ and 26.5 for NGC~6121 and NGC~6397, respectively. Additional deep observations are thus paramount to completely exploit these \jwst data.

\section{Electronic material}\label{sec:release}

We make our astro-photometric catalogs and stacked images publicly available. We release one catalog for each cluster that contains: equatorial coordinates in ICRS at the epoch of the \jwst observations;  $x$ and $y$ \jwst master-frame positions (in pixel; pixel scale of 40 mas pixel$^{-1}$); ID number of the source; photometric information for each filter; membership flag (flag$=$1 for cluster members and 0 otherwise). For NGC~6121, we also include a flag to discriminate between objects measured in the Northern (flag$=$1) and Southern (flag$=$2) fields.

For each filter, we provide: the calibrated VEGA magnitude corrected for differential reddening and for the input-output correction (Appendix~\ref{sec:artstar}); the raw instrumental magnitude; saturation flag (flag$=$0 for unsaturated objects and 9 otherwise); magnitude rms; quality-of-PSF-fit (QFIT) parameter; fractional flux within the fitting radius prior to neighbor subtraction; number of exposures in which a source was found; number of exposures used to measure the flux of a source; \radxs value; sky value; and sky rms.
    
We emphasize the following:
\begin{itemize}
    \item We included all objects detected in the \jwst/NIRISS images and provide ancillary information from the \hst data only for the sources found in common.
    \item We give a PM-based membership flag because improved PMs will be computed with state-of-the-art techniques \citep{2014BelliniPM,2022LibralatoPMcat} in a subsequent paper of the series.
    \item The 0.05-mag offset between the photometry in the Northern and Southern fields of NGC~6121 (Sect.\ref{sec:firstlook6121}) is included in the calibrated magnitude but not in the instrumental ones.
    \item All photometric quantities for a source are set to 0 if it is not measured in a given filter/camera/epoch.
    \item All photometric quantities but the magnitudes are set to 0 if a source is saturated.
    \item As described in Appendix~\ref{sec:artstar}, the input-output correction when the completeness level drops below $\sim$50\% is less constrained, so for stars fainter than this threshold, the correction was kept fixed at the value of the last reliable estimate. For NGC~6121, this threshold for each filter is: $m_{\rm F150W}^{\rm Northern}$$\sim$25.7, $m_{\rm F150W}^{\rm Southern}$$\sim$24.7, $m_{\rm F606W}$$\sim$27.2, $m_{\rm F814W}$$\sim$25.2. For NGC~6397, the threshold is set at $m_{\rm F150W}$$\sim$26.3, $m_{\rm F775W}$$\sim$26.4.
    \item The input-output correction for saturated stars included in the calibrated magnitudes is extrapolated from the unsaturated sources and it should not be considered in any analysis.
\end{itemize}

\end{appendix}

\label{LastPage}

\end{document}